\documentclass[letter,aps,prd,10pt,preprintnumbers,showkeys,superscriptaddress,nofootinbib,amsmath,amssymb,floatfix
]{revtex4-2}
\usepackage{times}
\usepackage{hyperref}
\usepackage{amsmath}
\usepackage{amssymb}
\usepackage{graphicx}
\usepackage{subfigure}
\usepackage{booktabs}
\usepackage{rotating}
\usepackage{longtable}
\usepackage{bm}
\usepackage{float}
\usepackage{epstopdf}
\usepackage{xcolor}
\usepackage{comment}

\makeatletter

\newcommand{\Rmnum}[1]{\expandafter\@slowromancap\romannumeral #1@}
\makeatother

\makeatletter
\newcommand*{\rom}[1]{\expandafter\@slowromancap\romannumeral #1@}
\makeatother
\begin{document}
\title{Rotating Black Holes Surrounded by Massive Vector Fields in Kaluza--Klein Gravity
}

 % % % % % %
\author{Farokhnaz Hosseinifar}
\email{f.hoseinifar94@gmail.com}
\affiliation{Center for Theoretical Physics, Khazar University, 41 Mehseti Str., Baku, AZ-1096, Azerbaijan}
 % % % % % %
\author{Shahin Mamedov}
\email{ctp@khazar.org}
\affiliation{Center for Theoretical Physics, Khazar University, 41 Mehseti Str., Baku, AZ-1096, Azerbaijan}
\affiliation{Institute for Physical Problems, Baku State University, Z.Khalilov 23, Baku, AZ-1148, Azerbaijan}
\affiliation{Institute of Physics, Ministry of Science and Education, H.Javid 33, Baku, AZ-1143, Azerbaijan}
 % % % % % %
\author{Kuantay Boshkayev}
\email{kuantay@mail.ru}
\affiliation{National Nanotechnology Laboratory of Open Type, 050040 Almaty, Kazakhstan}
\affiliation{Al-Farabi Kazakh National University, Al-Farabi ave. 71, 050040 Almaty, Kazakhstan}
 % % % % % %
\author{Soroush Zare}
\email{soroushzrg@gmail.com}
\affiliation{Helsinki Institute of Physics, University of Helsinki, P.O. Box 64, 00014 Helsinki, Finland}
 % % % % % %
\author{Filip Studni{\v{c}}ka}
\email{filip.studnicka@uhk.cz}
\affiliation{Department   of   Physics, Faculty of Science,  University   of   Hradec   Kr\'{a}lov\'{e}, Rokitansk\'{e}ho   62, 500   03   Hradec   Kr\'{a}lov\'{e},   Czechia}
 % % % % % %
\author{Hassan  Hassanabadi}
\email{hha1349@gmail.com}
\affiliation{Department   of   Physics, Faculty of Science,   University   of   Hradec   Kr\'{a}lov\'{e},  Rokitansk\'{e}ho 62, 500   03   Hradec   Kr\'{a}lov\'{e},   Czechia}
\affiliation{Khazar University, Department of Physics and Electronics, 41 Mahsati Str, AZ1096, Baku, Azerbaijan}
 % % % % % %

%%%%%%%%%%%%%%%%%%%%%%%%%%%%%%%%%%%%%%%%%%%%%%%%%%%%
%%%%%%%%%%%%%%%%%%%%%%%%%%%%%%%%%%%%%%%%%%%%%%%%%%%%
\begin{abstract}
In this paper, we introduce a rotating Kaluza--Klein (KK) black hole characterized by a massive vector field and a scalar field. We begin by identifying the horizons and mapping the allowed parameter space to differentiate black hole solutions from naked singularities. The thermodynamic analysis shows a phase transition by examining Hawking temperature and heat capacity. We also conduct a topological study of the thermodynamic potentials. The Hawking temperature indicates a conventional critical point, while the off--shell generalized free energy classifies the system into a specific universal group. We further investigate the geometry of the ergosphere and how it relates to the black hole's spin. Additionally, we look at astrophysical signs, such as the black hole shadow and the features of the thin accretion disk. Our results indicate that while the extra--dimensional changes significantly shift phase transition points and modify the shadow size, the essential topological class remains stable. This study provides a solid framework for distinguishing higher-dimensional gravity models through both thermodynamic and observational signs.
\end{abstract}
\maketitle
%%%%%%%%%%%%%%%%%%%%%%%%%%%%%%%%%%%%%%%%%%%%%%%%%%%%
%%%%%%%%%%%%%%%%%%%%%%%%%%%%%%%%%%%%%%%%%%%%%%%%%%%%
\section{Introduction}\label{Sec1}
%%%%%%%%%%%%%%%%%%%%%%%%%%%%%%%%%%%%%%%%%%%%%%%%%%%%
%%%%%%%%%%%%%%%%%%%%%%%%%%%%%%%%%%%%%%%%%%%%%%%%%%%%
The prediction of black holes is one of the greatest achievements of Einstein’s General Relativity (GR) \cite{einstein1916grundlagen,schwarzschild1916gravitationsfeld}. Gravity has been precisely defined by GR as the bending of four-dimensional spacetime for many years \cite{misner1973gravitation,wald2010general}.
The discovery of the Kerr solution, which offered a mathematical explanation for rotating black holes, was a turning point in this field \cite{kerr1963gravitational}.
Combining gravity with other fundamental forces, particularly electromagnetism, has proven to be a significant challenge in theoretical physics \cite{appelquist1987modern}.

Theodor Kaluza added a fifth spatial dimension in 1919, which was a revolutionary concept \cite{kaluza1921unitatsproblem}. He showed that a higher-dimensional framework could naturally include both Einstein’s gravity and Maxwell’s electromagnetism. He demonstrated how Maxwell's electromagnetism and Einstein's gravity could coexist peacefully in a higher dimensional framework. In 1926, Oskar Klein developed this idea further by providing a physical justification for the invisible nature of this additional dimension \cite{klein1926quantentheorie}. Klein proposed that the fifth dimension is compactified into a very small circle, with a radius typically at the scale of the Planck length \cite{salam1982kaluza,freund1982kaluza,gibbons1986black,bailin1987kaluza,duff1994kaluza,overduin1997kaluza,mashhoon1998dynamics}. This makes it undetectable at larger scales.

Early models of gravitational accretion were created by Hoyle, Lyttleton, and Bondi in the 1940s and 1950s to explain how stars grow \cite{hoyle1939effect,hoyle1940accretion,bondi1944mechanism,bondi1952spherically}. However, a paradigm shift occurred in the 1960s with the discovery of quasars.
To account for these extremely bright and compact energy sources, the focus changed from simple infall to the dynamics of rotating matter. It became clear that, because of angular momentum conservation, gas falling onto a compact object would ultimately create a flattened, radiating accretion disk \cite{lynden1969galactic}.
The foundational theoretical models for these thin disks, developed by Shakura and Sunyaev, as well as by Novikov and Thorne \cite{novikov1973astrophysics,page1974disk}, established how viscous processes could effectively turn gravitational energy into the huge radiation seen from these cosmic engines.
Since then, examining the unique thermal spectrum and radiation flux from accretion disks has become a key method for uncovering the properties of the central black hole, particularly its mass and spin \cite{salahshoor2018circular,zhu2020x,nozari2020quantum,heydari2020thin,mirzaev2023observational,wu2024thin,boshkayev2024luminosity,nieto2025accretion,koam2025probing,zare2026schwarzschild}.

Beyond the conventional thermodynamic description of black holes in terms of their temperature and entropy, a strong alternative view has emerged, treating them as topological defects in a vector field \cite{wei2022black,wei2015insight}.
The use of topological methods in physics has a rich history; they have been famously applied to classify solitons and monopoles in field theory \cite{mermin1979topological}. A significant development for gravitational physics was the framework established by Duan and his collaborators \cite{duan1993topological,duan1998topological}. They used a topological current theory to assign an integer--valued topological charge to a given solution. This topological charge provides a robust method to classify different black hole solutions into universal topological classes \cite{wei2022black}.
This classification has deep implications for black hole stability because different topological classes cannot be smoothly transformed into one another. Consequently, calculating this charge has become an essential tool for understanding the fundamental structure and robustness of new black hole solutions \cite{yerra2022topology,gogoi2023topology,wu2023topological,fang2023revisiting,fan2023topological,sadeghi2023bardeen,liu2023topological,dong2025thermodynamic,zhang2025new,wu2025novel,sekhmani2025lorentz}.

In modern KK frameworks, reducing spacetime dimensions creates new physical entities, like scalar and massive vector fields \cite{overduin1997kaluza,randall1999large,rizzo2001probes,zwiebach2004first}. Recently, Jusufi et al. \cite{jusufi2025black} examined a static black hole solution in this setting. Their study offered a detailed look at the metric derivation and properties of the accretion disk. They also investigated the black hole shadow and calculated the quasinormal modes.

However, most astrophysical black holes are defined by their parameters, so a static description is not enough to fully understand these objects. In this paper, we expand the study of such black holes to consider the rotating case. By using a method similar to Kerr-Newman geometry \cite{newman1965note}, we derive a new rotating metric. This strategy helps us explore how the combined effects of intrinsic spin and the fields from extra dimensions shape the curvature of spacetime and the observable characteristics of the black hole, including its shadow and accretion disk.

The remainder of this paper is structured to build our analysis from the foundational metric to its observable consequences. We begin in Sec. \ref{Sec2} by constructing our rotating KK metric and delineating the parameter space required for the existence of the event horizon.
In Sec. \ref{Sec5}, we examine the geometry of the ergosphere and quantify the frame-dragging effect, a key signature of rotation.
Then, we turn to the black hole fundamental properties: Sec. \ref{Sec3} is dedicated to its thermodynamics, where we calculate the Hawking temperature and investigate the possibility of black hole remnants. This is complemented in Sec. \ref{Sec4} by a topological analysis of the black hole's thermodynamic potentials, allowing for its classification into universal topological classes. Next, we explore the phenomenological and observational implications of our solution. We then model two primary observational signatures: the black hole shadow is calculated in Sec. \ref{Sec6}, and the dynamics and radiative properties of a thin accretion disk are analyzed in Sec. \ref{Sec7}, with a focus on identifying unique signatures from the extra dimension. Finally, we summarize our key findings and discuss potential avenues for future research in Sec. \ref{Sec12}.

%The structure of this paper is organized as follows: In Section \ref{Sec2}, we present the rotating KK metric and derive the allowed parameter space for the existence of horizons. Section \ref{Sec3} focuses on the thermodynamic properties, including Hawking temperature and black hole remnants. In Section \ref{Sec4}, we perform a topological analysis of the thermodynamic potentials to classify the black hole into universal groups. Section \ref{Sec5} examines the ergosphere and the frame-dragging effect. The black hole shadow is studied in Section \ref{Sec6}. In Section \ref{Sec7}, we analyze the dynamics and radiative properties of the thin accretion disk. Finally, we summarize our findings and provide concluding remarks in Section \ref{Sec12}.
%%%%%%%%%%%%%%%%%%%%%%%%%%%%%%%%%%%%%%%%%%%%%%%%%%%%
%%%%%%%%%%%%%%%%%%%%%%%%%%%%%%%%%%%%%%%%%%%%%%%%%%%%
\section{Derivation of the Rotating Kaluza--Klein Metric in the Kerr--Newman Framework}\label{Sec2}
%%%%%%%%%%%%%%%%%%%%%%%%%%%%%%%%%%%%%%%%%%%%%%%%%%%%
%%%%%%%%%%%%%%%%%%%%%%%%%%%%%%%%%%%%%%%%%%%%%%%%%%%%
In this section, we construct a new rotating black hole solution within the framework of the 5D KK theory. Our approach is to apply the Newman-Janis algorithm to a known static solution, which we use as a seed metric. Following the work of Jusufi et al. \cite{jusufi2025black}, we begin with the static spherically symmetric black hole solution given by
\begin{equation}\label{ds}
d s^2 = - f(r) d t^2 +\frac{1}{f(r)} dr^2+r^2(d\theta^2+\sin^2\theta\,d\phi^2).
\end{equation}
where $f(r)$ is defined as
\begin{eqnarray}
f(r)=1-\frac{2 M}{r}+\frac{\gamma  M^2 e^{-\frac{2 r}{\lambda }}}{r^2\lambda }(r+\lambda).
\end{eqnarray}
$M$ is the black hole's mass in the lapse function, and the parameters $\gamma$ and $\lambda$ represent the new physics derived from the KK framework. In particular, the mass of the spin--1 vector field that results from the dimensional reduction process is linked to the parameter $\gamma$, which models its impact on the spacetime geometry. In addition to adding Yukawa--type corrections to the standard Newtonian potential, the parameter $\lambda$ serves as a coupling constant that controls the strength of the interaction between this massive vector field and the black hole \cite{jusufi2025black}. 
%Collectively, these parameters, particularly $\lambda$, endow the solution with an effective charge, causing the black hole to exhibit behavior analogous to a charged RN or Kerr--Newman black hole, even in the absence of a true electric charge.

To introduce rotation, we employ the Newman--Janis algorithm \cite{newman1965note}, that transforms a static metric into a stationary, axisymmetric one.
The process begins by transforming the metric into coordinates $(u, r, \theta, \phi)$ and expressing it in a null tetrad basis. The core of the algorithm is a complexification of the coordinate space, where the radial coordinate $r$ and the null time coordinate $u$ are subjected to the transformation
\begin{equation}
	\begin{aligned}
		r &\rightarrow r' = r + ia\cos\theta,\qquad
		u &\rightarrow u' = u - ia\cos\theta,
	\end{aligned}
\end{equation}
with $a$ being the spin parameter. Finally, a coordinate transformation is applied to return to the $(t, r, \theta, \phi)$ coordinates, yielding the desired rotating metric. By applying this procedure to the seed metric \eqref{ds}, we derive our new rotating KK black hole solution. The line element is given by:
\begin{equation}\label{ds2}
	\begin{aligned}
		d s^2 =& -\left(\frac{\Delta-a^2 \sin^2\theta}{\Sigma}\right) d t^2 +\frac{\Sigma}{\Delta}d r^2+\Sigma d\theta^2-2a \sin^2\theta\left(1-\frac{\Delta-a^2 \sin^2\theta}{\Sigma}\right) d\theta d\phi
		\\&
		+\frac{\sin^2\theta}{\Sigma}\left((r^2+a^2)^2-\Delta a^2 \sin^2\theta\right)d \phi^2
	\end{aligned}
\end{equation}
where $\Delta$ and $\Sigma$ are defined as
\begin{equation}\label{delta}
	\Delta=a^2+r^2-2 M r+\frac{\gamma  M^2 e^{-\frac{2 r}{\lambda }} }{\lambda }(r+\lambda),\qquad\Sigma=r^2+a^2 \cos^2\theta.
\end{equation}
This metric describes a stationary, axisymmetric spacetime.
In the non--rotating limit where $a \to 0$, $\Sigma \to r^2$ and $\Delta \to r^2 f(r)$. Substituting these into the metric \eqref{ds2} correctly recovers the original static solution \eqref{ds}. Furthermore, the function $\Delta$ becomes $r^2 - 2Mr + a^2$ in the absence of the KK contribution, and our solution appropriately reduces to the standard Kerr metric of GR.

\begin{figure}[ht!]
	%\centering
	%   \begin{subfigure}[b]{0.5\textwidth}
		%   \centering
		\includegraphics[width=5.7cm]{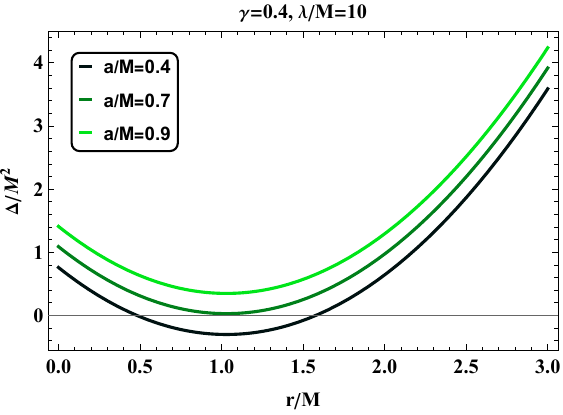} \hspace{0.2cm}
		\includegraphics[width=5.7cm]{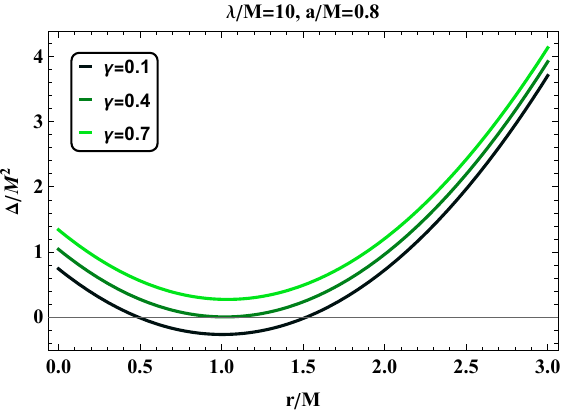} \hspace{0.2cm}
		\includegraphics[width=5.7cm]{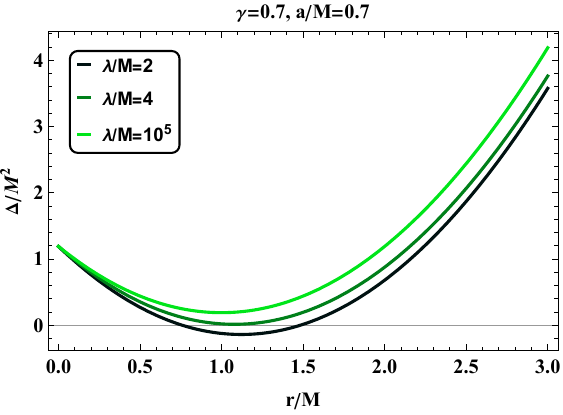} \hspace{-0.2cm}
		%\caption{}
		%    \end{subfigure}%
	\caption{The behavior of the function $\Delta/M^2$ in terms of $r/M$ for different choices of the spin and KK parameters.}\label{fig:delta}
\end{figure}

\begin{figure}[ht!]
	%\centering
	%   \begin{subfigure}[b]{0.5\textwidth}
		%   \centering
		\includegraphics[width=5.7cm]{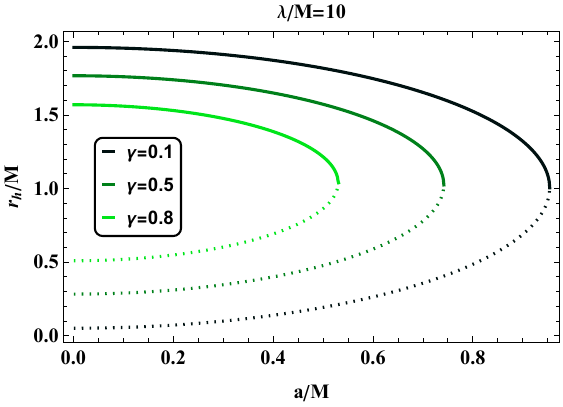} \hspace{0.5cm}
		\includegraphics[width=5.7cm]{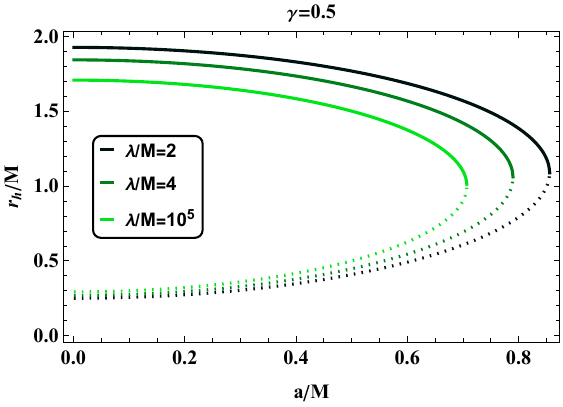} \hspace{-0.2cm}\\
		%\caption{}
		%    \end{subfigure}%
	\caption{Roots of $\Delta=0$ by varying $a/M$ and different selections of KK parameters. Solid lines represent $r_+$ and dotted lines indicate $r_-$.}\label{fig:horizon}
\end{figure}

\begin{figure}[ht!]
	\centering
	%   \begin{subfigure}[b]{0.5\textwidth}
		%   \centering
		\includegraphics[width=5.7cm]{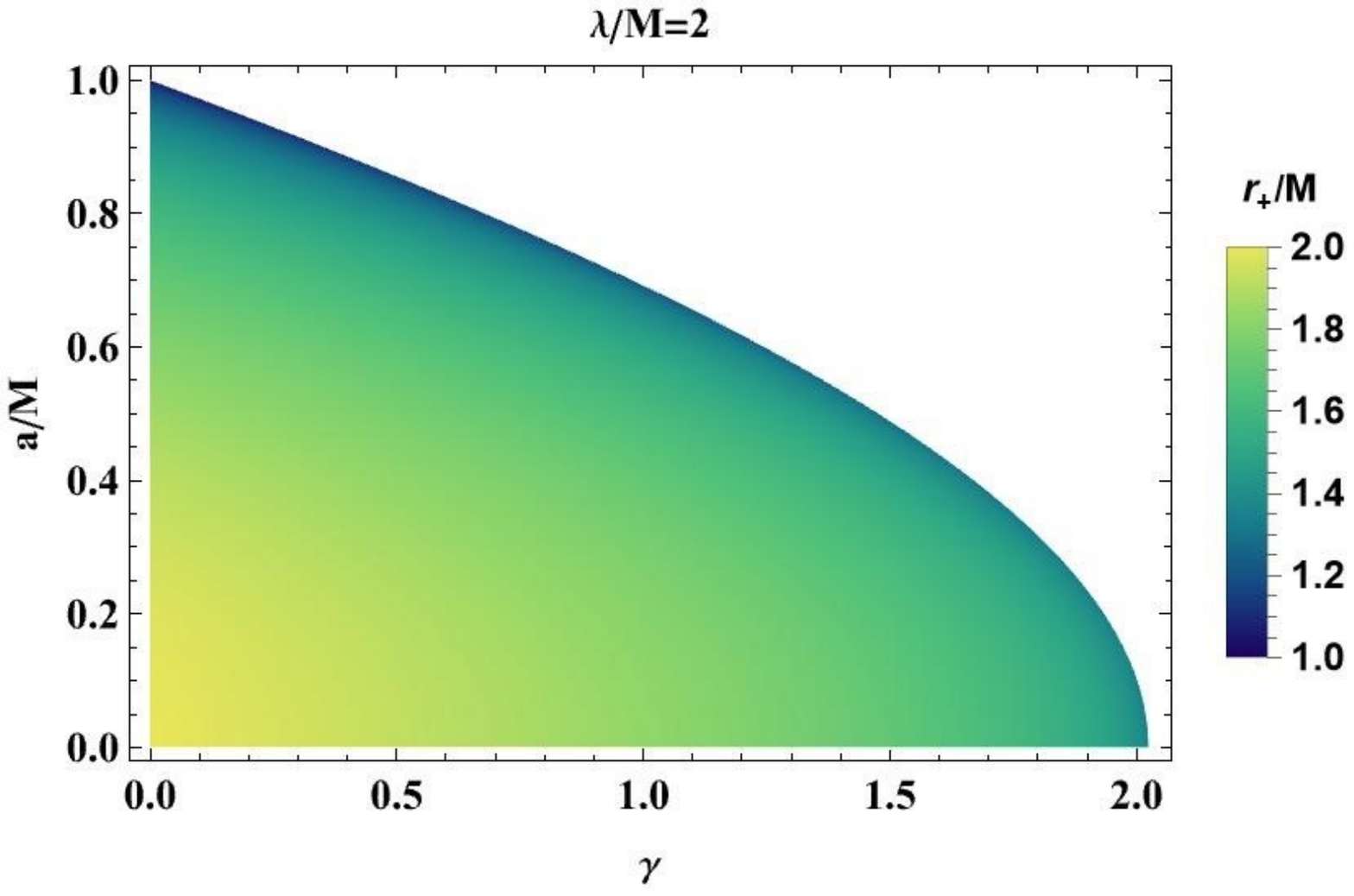} \hspace{0.2cm}
		\includegraphics[width=5.7cm]{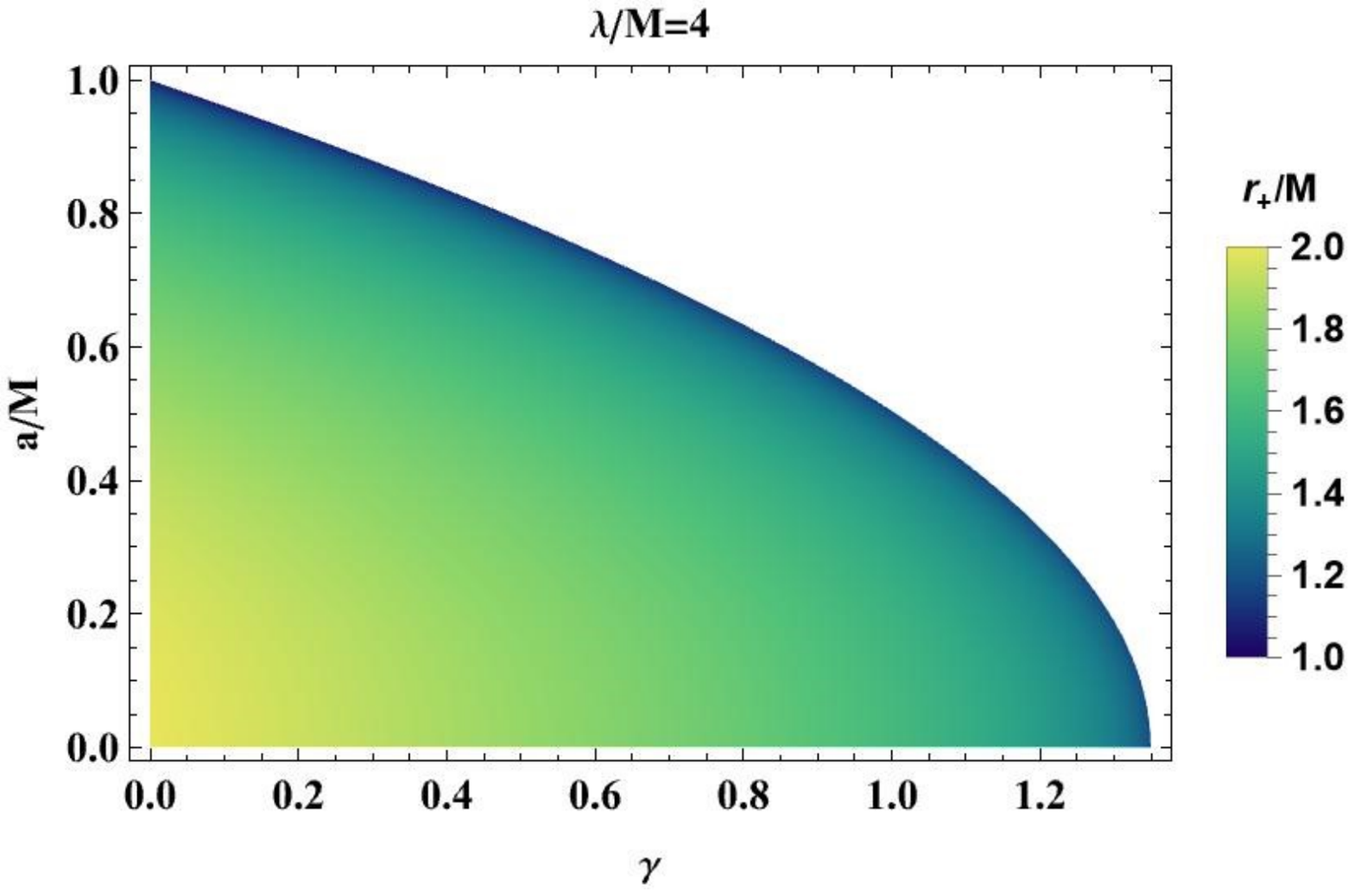} \hspace{0.2cm}
		\includegraphics[width=5.7cm]{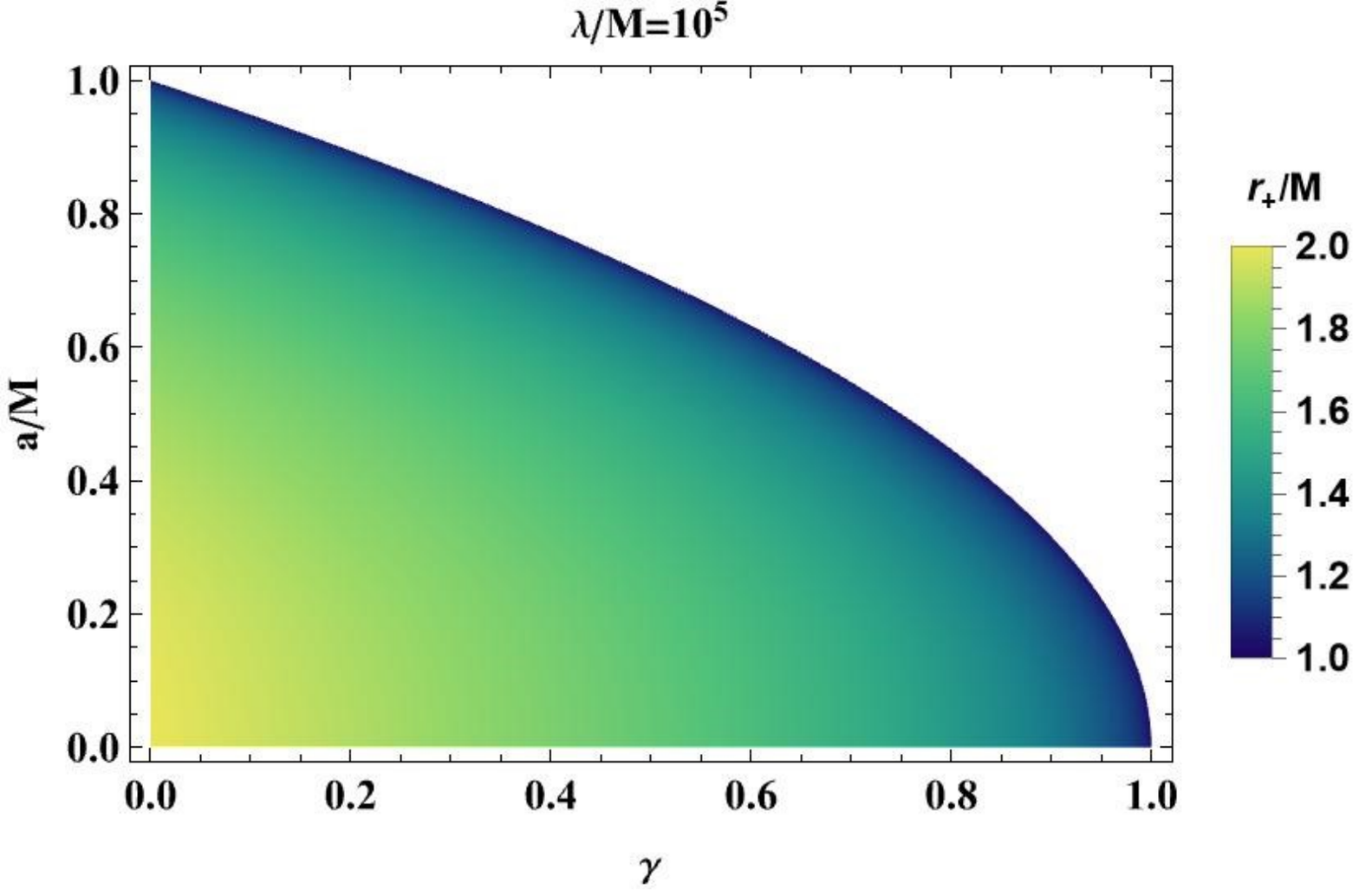} \hspace{-0.2cm}\\
		%\caption{}
		%    \end{subfigure}%
	\caption{Variations of event horizon radius in terms of spin and $\gamma$ for three cases of parameter $\lambda/M$. The magnitude of event horizon radius for each coordinates is specified by its color. The boundary of the colored and white regions represents the case where $r_+$ and $r_-$ merge together and $\Delta$ has only one root, and in the white regions $\Delta$ becomes horizonless.}\label{fig:Bhno}
\end{figure}

Now, we examine the horizon structure of the rotating KK black hole $(g^{rr}\big|_{r=r_h}=0)$ by studying the roots of the metric function $\Delta(r)$. The horizons are found from $\Delta\big|_{r=r_h}=0$. Figure \ref{fig:delta} displays the behavior of $\Delta/M^2$ by changing one parameter while keeping the others fixed.
Our findings reveal that depending on the spin parameter $a/M$, $\gamma$, and $\lambda/M$, the black hole can display three different configurations. $\Delta$ may consists of two solutions: A Cauchy horizon $r_-$ and an event horizon $r_+$.
For some selections there is an extremal black hole, where the two horizons merge into a single root $r_-=r_+$ at certain critical values of the parameters. Also, when the parameters exceed critical limits, $\Delta(r)$ has no real roots, and the event horizon disappears. We conduct our analysis in a regime where $\Delta$ has two root. Figure \ref{fig:horizon} illustrates the variation of $\Delta(r)$ roots in terms of $a/M$.
These figures show how the roots change as a function of the spin for different values of $\gamma$ with a fixed $\lambda/M$ and vice versa. It is clear that while all parameters affect the horizon locations, the effect of $\lambda/M$ is much smaller compared to $a/M$ and $\gamma$. Additionally, the influence of $\lambda/M$ is more noticeable at smaller $\lambda/M$, but its effect lessens at larger value of this parameter.
Figure \ref{fig:Bhno} demonstrates the region where $\Delta$ has roots as a function of the parameters $\gamma$ and $a/M$ for three different cases of $\lambda/M$. 
In this figure, the colored region represents the parameter space that allows for the existence of the event horizon, corresponding to the condition where $\Delta=0$ has real roots. The color at each point indicates the value of the event horizon radius for that specific set of parameters. The boundary separating the colored and white regions (horizonless areas) corresponds to the extremal limit, where the Cauchy and event horizons merge. Furthermore, consistent with the observations from Figs. \ref{fig:delta} and \ref{fig:horizon}, increasing any one of the parameters $(a/M,\,\gamma,\,\lambda/M)$ while holding the others fixed leads to a decrease in the event horizon radius. Consequently, this diminishes the allowed parameter space for the remaining variables within which a black hole with an event horizon can exist.
Additionally, a key observation is that the sensitivity of the horizon radii to the parameter $\lambda/M$ is highly non--linear. The transition from $\lambda/M = 2$ to $\lambda/M = 4$ produces a significantly more prominent shift in the horizon locations than the transition between much larger values, such as $\lambda/M = 100000$. This suggests that the influence of the parameter $\lambda$ saturates at higher values, whereas its impact is most critical in the small--parameter regime.
%%%%%%%%%%%%%%%%%%%%%%%%%%%%%%%%%%%%%%%%%%%%%%%%%%%%
%%%%%%%%%%%%%%%%%%%%%%%%%%%%%%%%%%%%%%%%%%%%%%%%%%%%
\section{Influence of Kaluza--Klein fields on the Ergosphere}\label{Sec5}
%%%%%%%%%%%%%%%%%%%%%%%%%%%%%%%%%%%%%%%%%%%%%%%%%%%%
%%%%%%%%%%%%%%%%%%%%%%%%%%%%%%%%%%%%%%%%%%%%%%%%%%%%
The ergosphere is one of the most distinctive regions of a rotating black hole spacetime, arising from the strong frame--dragging induced by rotation. In this region, frame dragging becomes so extreme that no observer can remain static with respect to asymptotic infinity. The ergosphere is bounded internally by the event horizon and externally by the static limit surface, where the Killing vector associated with stationarity becomes null. Because the geometry of the ergosphere is highly sensitive to the metric coefficients, it provides an important probe of modifications to General Relativity. In the present rotating KK framework, both the massive vector field and the scalar sector alter the spacetime structure, leading to nontrivial deformations of the ergoregion \cite{Penrose1971Extraction,Gariel2013Unbound,Blandford2022Ergomagnetosphere,Hassanabadi2025Ergosphere}.

The boundary of the static limit surface region is determined by the condition where the time--like Killing vector becomes null. Given our rotating KK black hole metric, $r_s$ is found from
\begin{equation}
	\Delta(r_s)=a^2 \sin^2\theta.
\end{equation}
Also, the event horizon is computed from $\Delta(r_+)=0$, thus, the ergosphere is sensitive to the angular coordinate. Specifically, the ergosphere coincides with $r_+$ at $\theta=0,\,\pi$ and reaches its maximum at $\theta=\pi/2$, and at the equatorial plane $\Delta(r_s)=a^2$.
Unlike the event horizon, the static limit surface depends explicitly on the polar angle $\theta$, implying that the ergosphere has an oblate geometry. The separation between the event horizon and the static limit surface is maximal on the equatorial plane and vanishes along the rotation axis. This angular dependence reflects the anisotropic nature of rotational frame dragging.
%The thickness of the ergosphere which is defined as the difference between the static limit surface and the event horizon is calculated from $\Delta r=r_s-r_+$.

The KK parameters $\gamma$ and $\lambda$ modify the structure of the ergosphere through their contributions to the metric function $\Delta(r)$. Since these parameters effectively alter the gravitational potential surrounding the black hole, they modify the location of both the event horizon and the static limit surface. In particular, increasing $\gamma$ tends to enlarge the separation between these surfaces near the equatorial plane, leading to a thicker ergoregion. The parameter $\lambda$ produces a comparatively weaker effect, consistent with its subdominant influence on the horizon structure discussed previously. These modifications indicate that extra--dimensional fields can influence rotational energy extraction processes occurring inside the ergosphere.

\begin{figure}[ht!]
\centering
	\includegraphics[width=5.6cm]{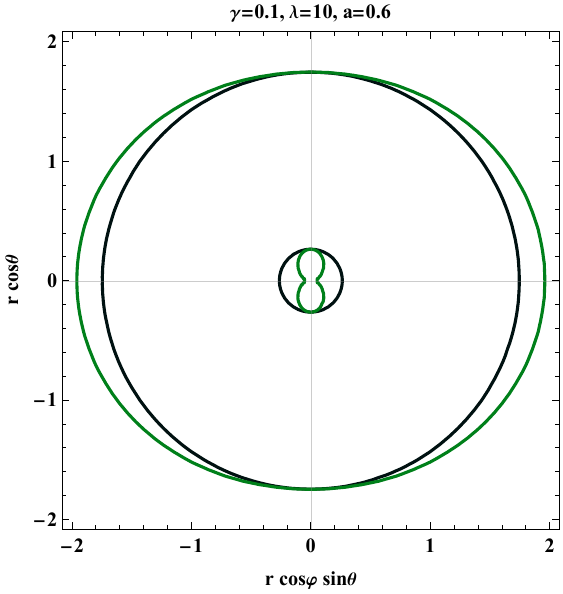} \hspace{0.2cm}
	\includegraphics[width=5.6cm]{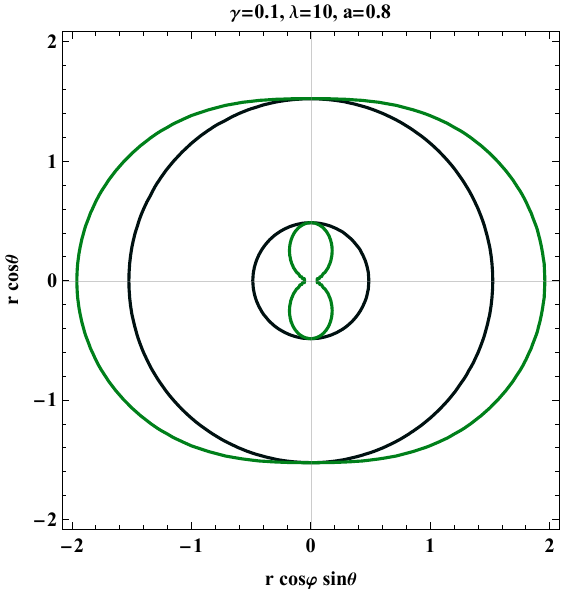} \hspace{0.2cm}
	\includegraphics[width=5.6cm]{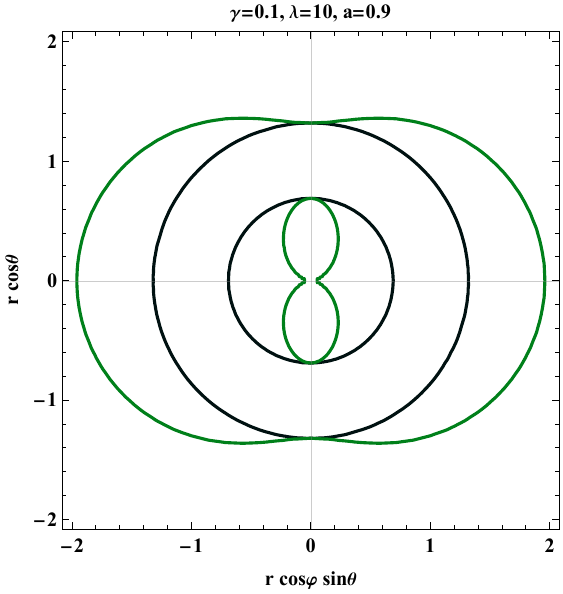} \hspace{-0.2cm}\\
	\includegraphics[width=5.6cm]{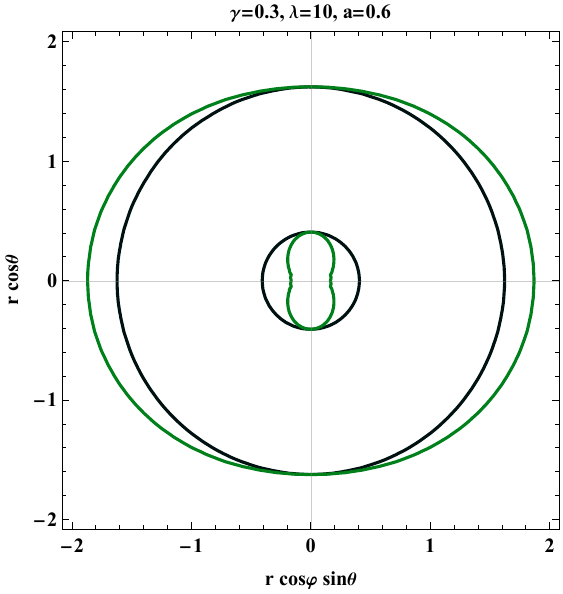} \hspace{0.2cm}
	\includegraphics[width=5.6cm]{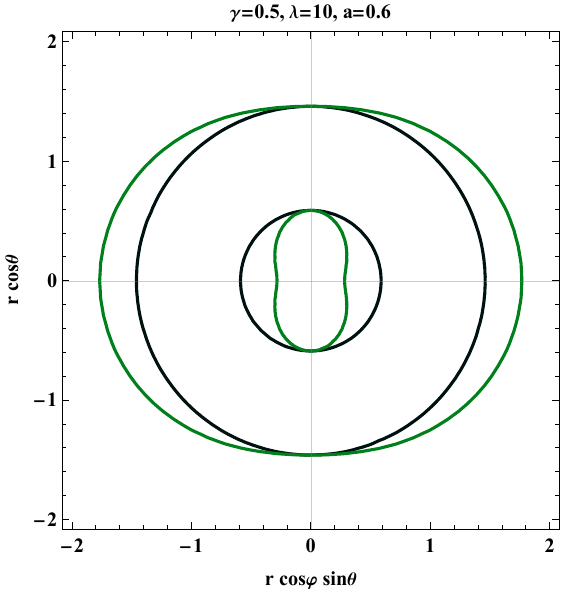} \hspace{0.2cm}
	\includegraphics[width=5.6cm]{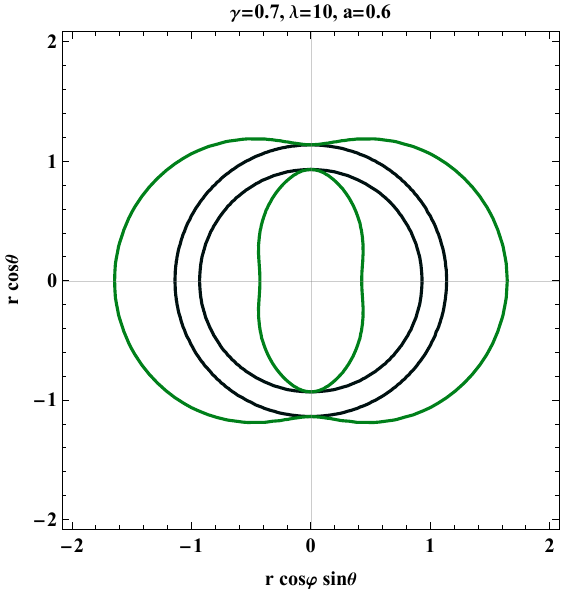} \hspace{-0.2cm}\\
	\includegraphics[width=5.6cm]{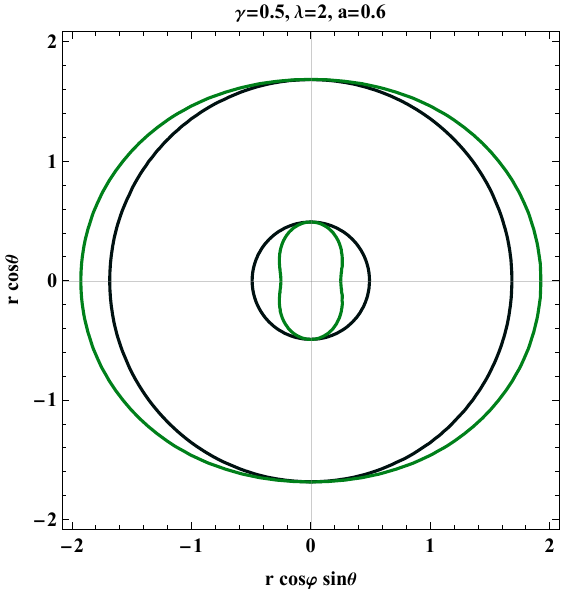} \hspace{0.2cm}
	\includegraphics[width=5.6cm]{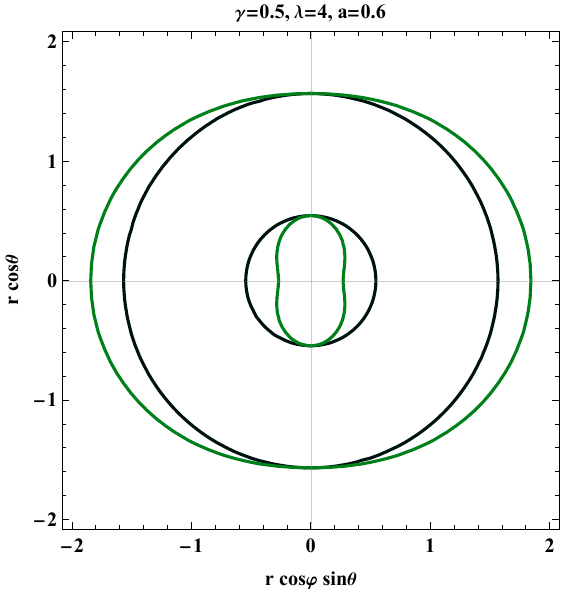} \hspace{0.2cm}
	\includegraphics[width=5.6cm]{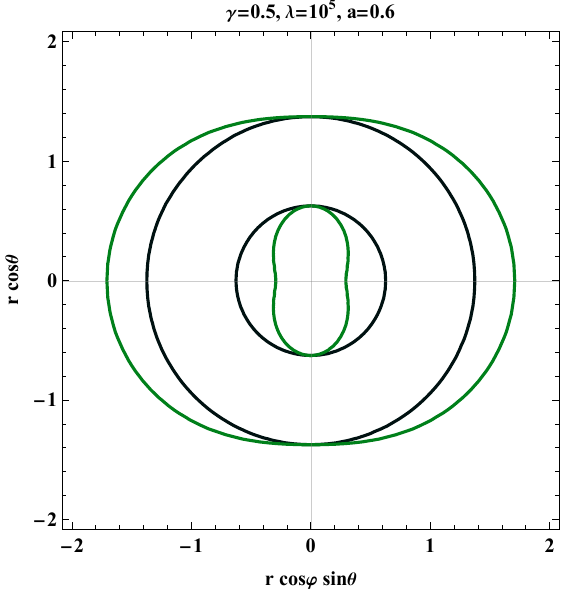} \hspace{-0.2cm}\\
\caption{The variation of the ergosphere in the X--Y plane for different selections of spin parameter (upper panels), $\gamma$ (middle panels), and $\lambda$ (lower panels). }\label{fig:Er}
\end{figure}

Figure \ref{fig:Er} illustrates the deformation of the ergosphere in the $X–Y$ plane for different choices of the spin and KK parameters.
The outer curve corresponds to the static limit surface, while the inner curve represents the event horizon. As expected, increasing the spin parameter significantly enlarges the ergosphere, particularly around the equatorial region. This behavior arises from the enhanced frame--dragging induced by rapid rotation, which pushes the static limit surface farther from the horizon.
The KK parameter $\gamma$ also produces noticeable modifications to the ergoregion. Larger values of $\gamma$ increase the distortion and thickness of the ergosphere, indicating that the extra–dimensional vector field strengthens rotational effects in the near--horizon geometry. In contrast, variations of $\lambda$ produce comparatively smaller changes, although they still lead to measurable shifts in the static limit surface for sufficiently small $\lambda$.
The enlargement of the ergosphere has important astrophysical implications because it directly affects mechanisms of rotational energy extraction, such as the Penrose process and superradiant scattering. A larger ergoregion generally enhances the efficiency of these processes by increasing the available region in which negative--energy states can exist.
Since the ergosphere allows particles to possess negative energies relative to infinity, rotational energy can be extracted from the black hole through the Penrose process. The modification of the ergoregion by the KK fields, therefore suggests that extra--dimensional effects may influence the efficiency of energy extraction and jet production near rotating compact objects. This may have observational consequences for active galactic nuclei and relativistic jet systems.
%%%%%%%%%%%%%%%%%%%%%%%%%%%%%%%%%%%%%%%%%%%%%%%%%%%%
%%%%%%%%%%%%%%%%%%%%%%%%%%%%%%%%%%%%%%%%%%%%%%%%%%%%
\subsection*{3.1.$\;\;\;$ Approximate Calculation of The Slow--Rotation Expansion of The Ergosphere Thickness\label{Sec5.1}}
%%%%%%%%%%%%%%%%%%%%%%%%%%%%%%%%%%%%%%%%%%%%%%%%%%%%
%%%%%%%%%%%%%%%%%%%%%%%%%%%%%%%%%%%%%%%%%%%%%%%%%%%%
In this section, we want to give a useful approximation for the slow--rotation expansion of the ergosphere thickness.\\
Using the definitions $\Delta$ (Eq. \eqref{delta}), with $r_+$ from $\Delta(r_+)=0$ and $r_s$ from $\Delta(r_s)=a^2\sin^2\theta$, the thickness $l(\theta)=r_s-r_+$.
\\For $a/M\ll 1$, let $r_0$ be the outer horizon of the corresponding nonrotating solution:
\begin{equation}
r_0^2- 2 M r_0+\frac{\gamma M^2}{\lambda}\exp^{(-2r_0/\lambda)}(r_0+\lambda)=0
\end{equation}
Meanwhile, we can define $\Delta(r)=F(r)+a^2$ and the event horizon satisfies $F(r_+)=-a^2$ and the static horizon satisfies $\Delta(r_s)=a^2\sin^2\theta$ or $F(r_s)+a^2=a^2\sin^2\theta$ or $F(r_s)=-a^2\cos^2\theta$.
\\Now, let $r_0$ be the nonrotating outer horizon, defined by $F(r_0)=0$. For small spin, the expansion of $F(r)$ around $r_0$ reads
\begin{equation}
F(r) \approx F(r_0)+F'(r_0)(r-r_0).
\end{equation}
Since $F(r_0)=0$, then $F(r)\approx F'(r_0)(r-r_0)$. Then, $r_+-r_0\approx -a^2/F'(r_0)$ and $r_s-r_0\approx -a^2 \cos^2\theta/F'(r_0)$.
\\By the above results, the ergosphere thickness is $l(\theta)=r_s-r_+\approx a^2\sin^2\theta/F'(r_0)$ and by using the definition of $F(r)$, we have
\begin{equation}
\frac{l(\theta)}{M}\approx\frac{(a/M)^2\sin^2\theta}{2 x_0-2-\gamma \exp^{(-2x_0/L)}(1/L+2x_0/L)},\qquad\text{where}\qquad x_0=r_0/M,\;L=\lambda/M
\end{equation}
The denominator of $F(r_0)$ is given by $F'(r_0)$, where $r_0$ denotes the outer horizon of the corresponding nonrotating solution. For a nonextremal black hole, the outer horizon corresponds to a simple root of $F(r)$, implying $F'(r_0)>0$. In the extremal limit, the two horizons merge and the root becomes degenerate, leading to $F'(r_0)=0$. Therefore, the approximation remains valid only away from extremality.
%%%%%%%%%%%%%%%%%%%%%%%%%%%%%%%%%%%%%%%%%%%%%%%%%%%%
%%%%%%%%%%%%%%%%%%%%%%%%%%%%%%%%%%%%%%%%%%%%%%%%%%%%
\section{Thermodynamics and Local Thermal Stability}\label{Sec3}
%%%%%%%%%%%%%%%%%%%%%%%%%%%%%%%%%%%%%%%%%%%%%%%%%%%%
%%%%%%%%%%%%%%%%%%%%%%%%%%%%%%%%%%%%%%%%%%%%%%%%%%%%
In this section, we investigate the key thermodynamic properties of the rotating KK black hole, including its mass, entropy, Hawking temperature, heat capacity, and generalized free energy.
Mass of the black hole as function of horizon radius is obtained from $\Delta(r_+)=0$ and reads
\begin{eqnarray}\label{Mplus}
M_+=\frac{e^{r_+/\lambda } \left(\lambda  r_+ e^{r_+/\lambda }-\sqrt{\lambda  \left(\lambda  r_+^2 e^{\frac{2 r_+}{\lambda }}-\gamma  \left(a^2+r_+^2\right) (r_+ +\lambda)\right)}\right)}{\gamma  (r_+ +\lambda )}
\end{eqnarray}
Using the horizon area relation, the entropy of the black hole employing Eq. \eqref{ds2} is obtained as \cite{jacobson1994black,hayward1999dynamic}
\begin{eqnarray}\label{Ent}
S_+=\pi  \left(a^2+r_+^2\right).
\end{eqnarray}
Although the above entropy is functionally equal to the entropy of the Kerr black hole, its numerical value is implicitly affected by the KK parameters. Because the event horizon of this black hole as indicated in Figures \ref{fig:delta} to \ref{fig:Bhno}, is dependent on the parameters $\gamma$ and $\lambda$ \cite{nozari2015black,saghafi2021thermodynamics}.

\begin{figure}[ht!]
%\centering
 %   \begin{subfigure}[b]{0.5\textwidth}
    \centering
  \includegraphics[width=5.7cm]{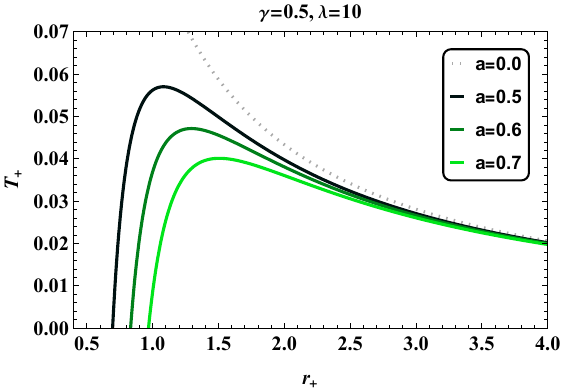} \hspace{0.2cm}
  \includegraphics[width=5.7cm]{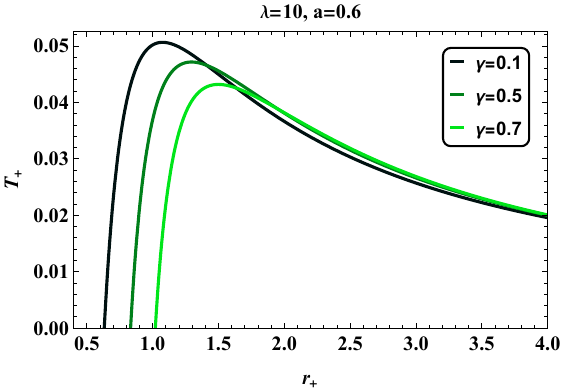} \hspace{0.2cm}
  \includegraphics[width=5.7cm]{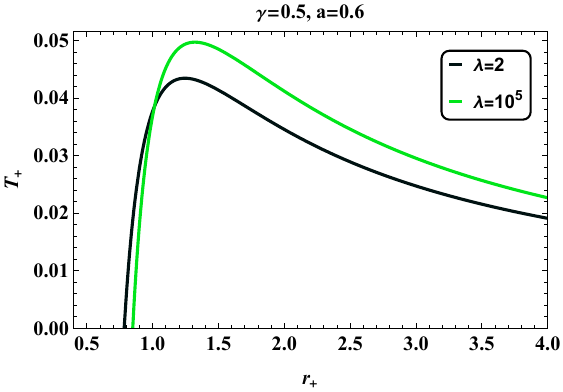} \hspace{-0.2cm}
%\caption{}
%    \end{subfigure}%
    \caption{The behavior of Hawking temperature as function of $r_+$ for different cases of spin and KK parameters. For the rotating KK black hole, Hawking temperature curve exhibit a maximum but, there is no extremum in the static case.}\label{fig:Temp}
\end{figure}

The Hawking temperature of the black hole employing Eqs. \eqref{Mplus} and \eqref{Ent}, is given by \cite{fursaev1995temperature,banerjee2008noncommutative}
\begin{equation}\label{Temp}
\begin{aligned}
T_+=&\frac{dM_+}{dS_+}\\
=&
%\\&
e^{\frac{r_+}{\lambda }}\frac{\gamma  (\lambda +r_+) \left(a^2 (\lambda +2 r_+)+r_+ \left(2 \lambda ^2+2 r_+^2+3 \lambda  r_+\right)\right)-2 \lambda  r_+ e^{\frac{2 r_+}{\lambda }} \left(\lambda ^2+2 r_+^2+2 \lambda  r_+\right)}{4 \pi  \gamma  r_+ (\lambda +r_+)^2 \sqrt{\lambda  \left(\lambda  r_+^2 e^{\frac{2 r_+}{\lambda }}-\gamma  \left(a^2+r_+^2\right) (\lambda +r_+)\right)}}
+e^{\frac{2 r_+}{\lambda }}\frac{\lambda ^2+2 r_+^2+2 \lambda  r_+}{2 \pi  \gamma  r_+ (\lambda +r_+)^2}.
\end{aligned}
\end{equation}
Figure \ref{fig:Temp} illustrates the variation of the Hawking temperature in terms of event horizon radius for different parameter sets.
The Hawking temperature for $a>0$ exhibit a maximum, which corresponds to a phase transition at $\partial_{r_+}T_+\big|_{r_+=r_c}=0$ during its evolution.
Increasing the spin parameter while keeping other parameters the same, causes the temperature to drop at a fixed radius. Additionally, the peak temperature decreases and shifts toward a larger horizon radius.
On the other hand, the effect of enhancing the parameter $\lambda$ on the temperature is not significant, but it becomes more apparent at larger horizon radii, resulting in a rise in temperature and shifting its peak to larger horizon radius.
For the parameter $\gamma$, an increase causes the maximum of temperature to occur at a larger radius with a lower value.
Interestingly, variations in KK parameters $\gamma$ and $\lambda$ have a dual effect: at small radii, increasing parameter $\gamma$ ($\lambda$) lowers (uppers) the temperature, while at larger radii, it raises it.

\begin{figure}[ht!]
%\centering
 %   \begin{subfigure}[b]{0.5\textwidth}
    \centering
  \includegraphics[width=5.7cm]{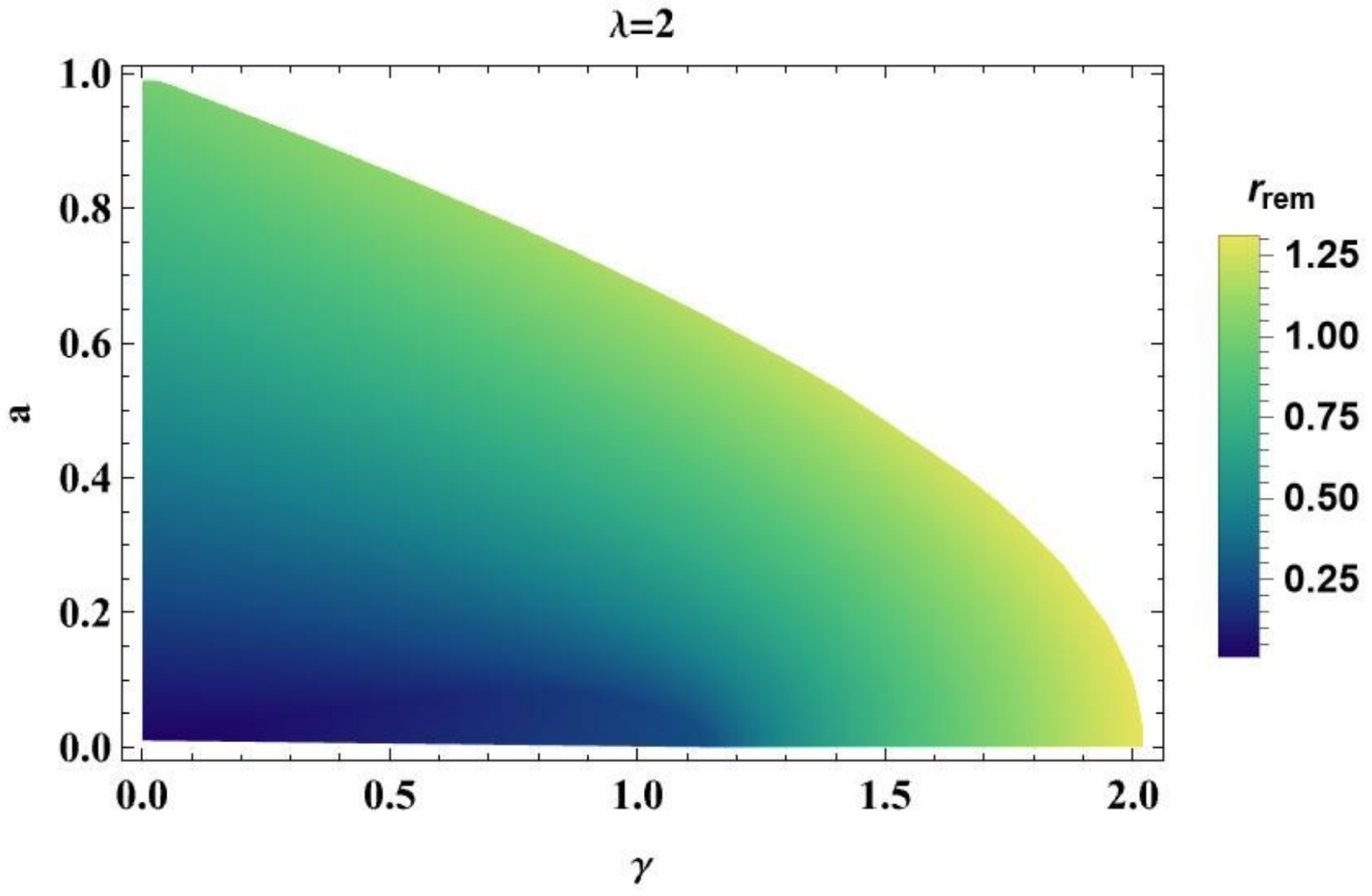} \hspace{0.2cm}
  \includegraphics[width=5.7cm]{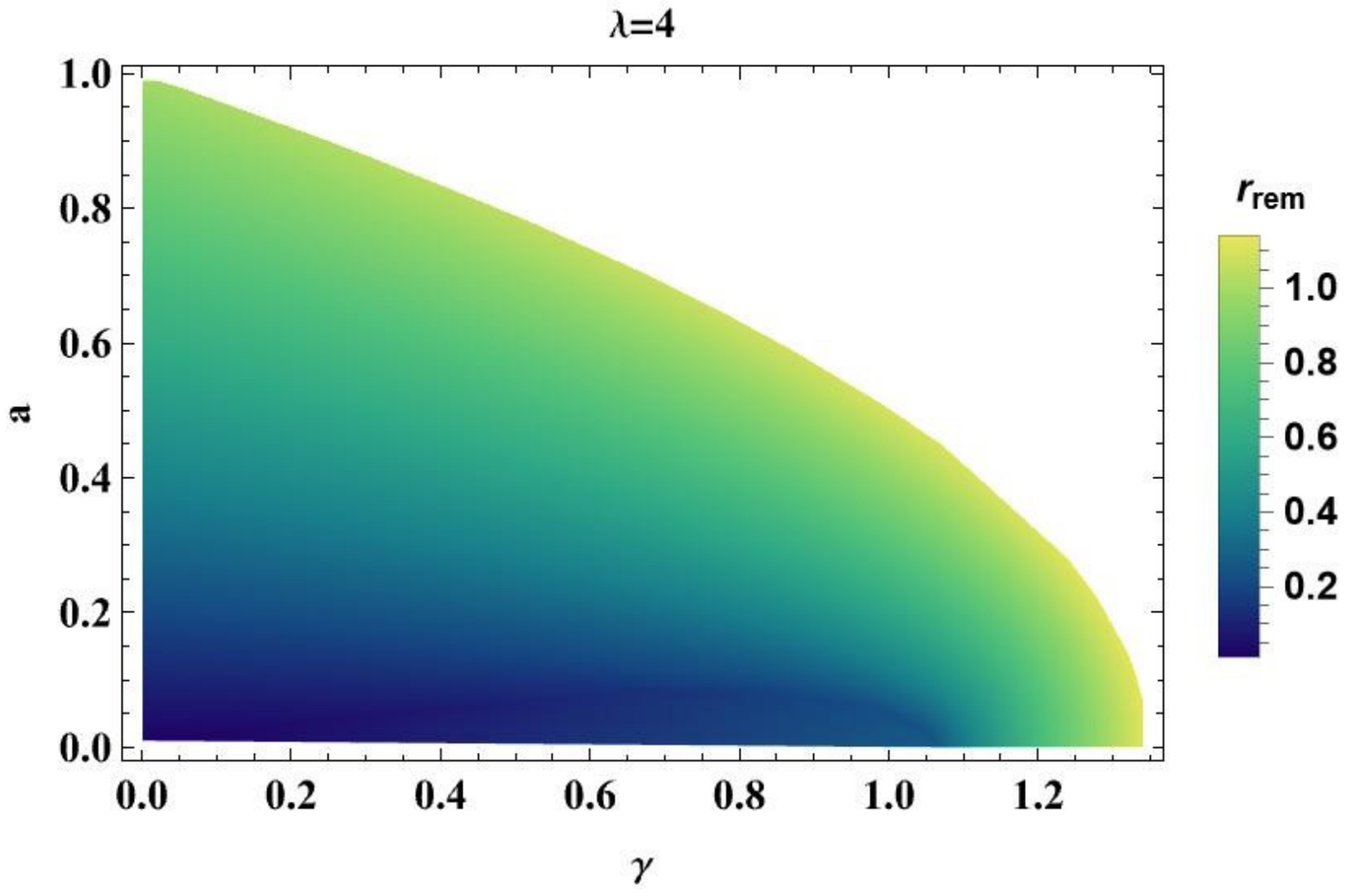} \hspace{0.2cm}
  \includegraphics[width=5.7cm]{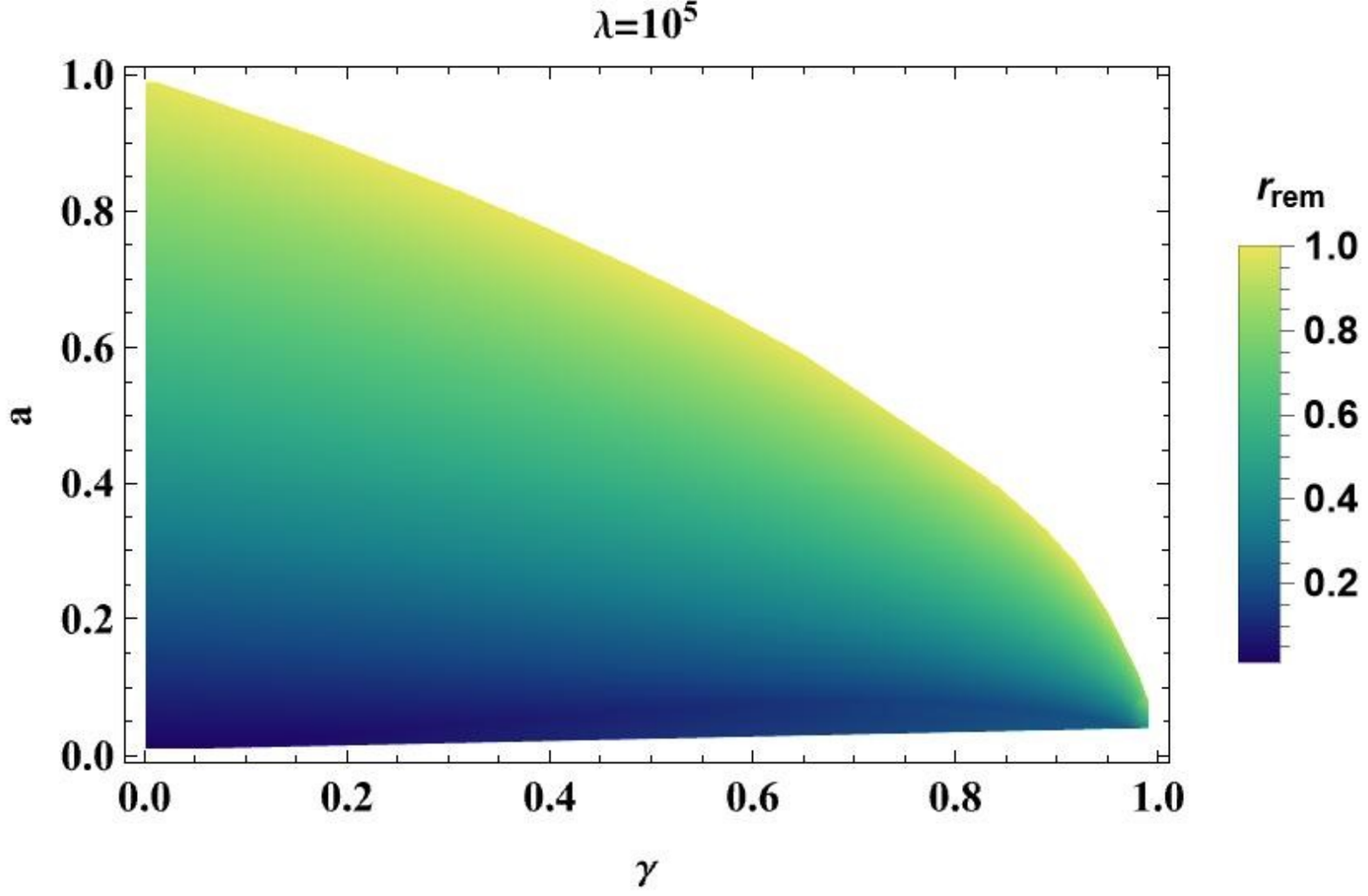} \hspace{-0.2cm}\\
%\caption{}
%    \end{subfigure}%
    \caption{Variation of the black hole remnant in terms of $a$ and $\gamma$ for three selection of $\lambda$. The magnitude of the remnant radius is represented by its color.}\label{fig:Rem}
\end{figure}

Another thermodynamic property studied in this section is the remnant radius, which is computed by finding the event horizon radius $r_+$ that satisfies \cite{gogoi2024quasinormal,zare2024influences}
\begin{eqnarray}
T_+\big|_{r_+=r_{rem}}=0,
\end{eqnarray}
and as anticipated from Figure \ref{fig:Temp}, although the absolute changes in the size of the black hole remnant are small, increasing either the parameter $\gamma$, $\lambda$ or $a$, while keeping other parameters constant, results in a larger remnant size. Figure \ref{fig:Rem}, displays the remnant radii behavior as a function of $a$ and $\gamma$ for three different cases of $\lambda$, represented over the parameter range where the event horizon exists.

\begin{figure}[ht!]
%\centering
 %   \begin{subfigure}[b]{0.5\textwidth}
 %   \centering
  \includegraphics[width=5.7cm]{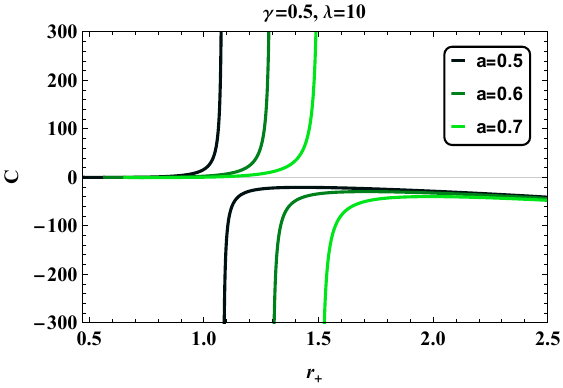} \hspace{0.2cm}
  \includegraphics[width=5.7cm]{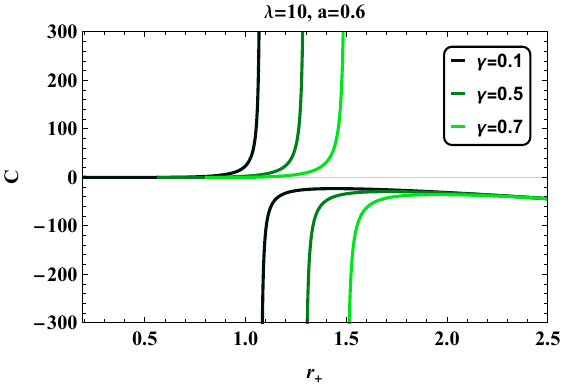} \hspace{0.2cm}
  \includegraphics[width=5.7cm]{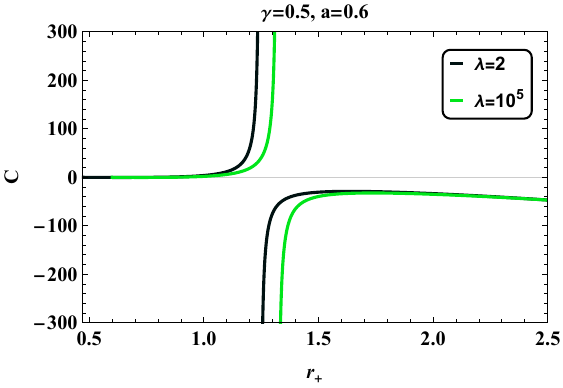} \hspace{-0.2cm}
%\caption{}
%    \end{subfigure}%
    \caption{Heat capacity of the rotating KK black hole as function of $r_+$ for various selections of spin and KK parameters.}\label{fig:Cp}
\end{figure}

To investigate the thermodynamic stability of the black hole, we calculate heat capacity using Eqs. \eqref{Ent} and \eqref{Temp} as \cite{brown1994temperature}
\begin{equation}
\begin{aligned}
C=\frac{\partial M_+}{\partial T_+}.
\end{aligned}
\end{equation}
Figure \ref{fig:Cp} illustrates the behavior of the heat capacity as a function of event horizon radius.
The heat capacity diverges at the a critical point $r_c$ where the Hawking temperature reaches its maximum, which is characteristic of a second--order phase transition. A positive heat capacity indicates that the black hole's temperature increases as it absorbs energy, whereas a negative heat capacity signifies that despite energy absorption, the black hole's temperature decreases. Therefore, a positive heat capacity corresponds to a thermodynamically stable state, while a negative heat capacity reflects its instability. Our results show that an increase in either the spin parameter or the KK parameters, while keeping other parameters fixed, shifts this phase transition point toward a larger horizon radius. This behavior is consistent with our previous observations of the Hawking temperature, where the peak temperature also shifted outward.

\begin{figure}[ht!]
%\centering
 %   \begin{subfigure}[b]{0.5\textwidth}
 %   \centering
  \includegraphics[width=5.7cm]{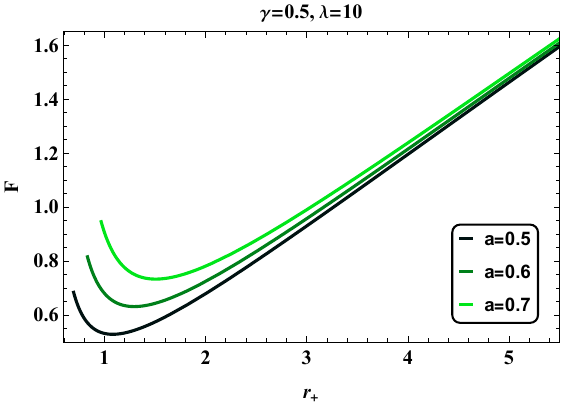} \hspace{0.2cm}
  \includegraphics[width=5.7cm]{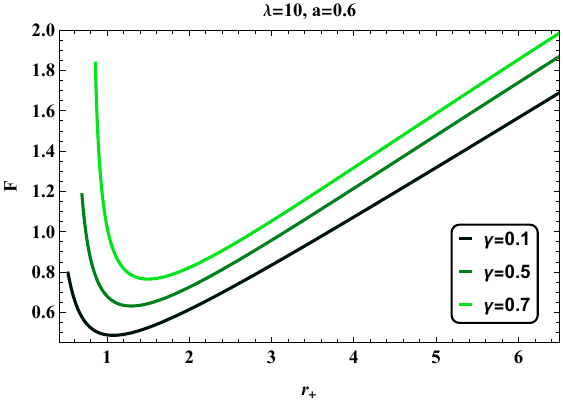} \hspace{0.2cm}
  \includegraphics[width=5.7cm]{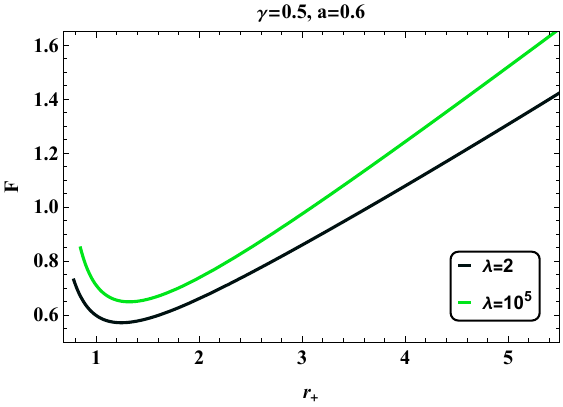} \hspace{-0.2cm}
%\caption{}
%    \end{subfigure}%
    \caption{Variations of generalized free energy in terms of $r_+$ considering different cases of the rotating KK black hole parameters.}\label{fig:OnFree}
\end{figure}

The final quantity investigated in this section is the generalized free energy, which governs the global stability of the black hole. The generalized free energy of the black hole employing Eqs. \eqref{Mplus}, \eqref{Ent}, and \eqref{Temp} is derived as \cite{kazakov2001free,li2020thermodynamics}
\begin{equation}
\begin{aligned}
F=& M_+-T_+ S_+\\
=& e^{\frac{r_+}{\lambda }}\frac{-r_+ \left(2 (\lambda +r_+) \sqrt{\lambda  \left(\lambda  r_+^2 e^{\frac{2 r_+}{\lambda }}-\gamma  \left(a^2+r_+^2\right) (\lambda +r_+)\right)}+r_+ e^{\frac{r_+}{\lambda }} \left(2 r_+^2-\lambda ^2\right)\right)-a^2 e^{\frac{r_+}{\lambda }} \left(\lambda ^2+2 r_+^2+2 \lambda  r_+\right)}{2 \gamma  r_+ (\lambda +r_+)^2}
\\&+
e^{\frac{r_+}{\lambda }}\left(a^2+r_+^2\right) \frac{2 \lambda  r_+ e^{\frac{2 r_+}{\lambda }} \left(\lambda ^2+2 r_+^2+2 \lambda  r_+\right)-\gamma  (\lambda +r_+) \left(a^2 (\lambda +2 r_+)+r_+ \left(2 \lambda ^2+2 r_+^2+3 \lambda  r_+\right)\right)}{4 \gamma  r_+ (\lambda +r_+)^2 \sqrt{\lambda  \left(\lambda  r_+^2 e^{\frac{2 r_+}{\lambda }}-\gamma  \left(a^2+r_+^2\right) (\lambda +r_+)\right)}}.
\end{aligned}
\end{equation}
Figure \ref{fig:OnFree} illustrates generalized free energy curves as function of event horizon radius.
The free energy profile displays a minimum, which serves as another indicator of the phase transition. This minimum corresponds precisely to the critical radius $r_c$ where the Hawking temperature reaches its maximum and the heat capacity diverges. It is evident that increasing either $a$ or KK parameters (at fixed values of other parameters), results in two key effects, first that the location of the transition shifts toward a larger radius, and second, the overall magnitude of the free energy increases.
%%%%%%%%%%%%%%%%%%%%%%%%%%%%%%%%%%%%%%%%%%%%%%%%%%%%
%%%%%%%%%%%%%%%%%%%%%%%%%%%%%%%%%%%%%%%%%%%%%%%%%%%%
\section{Topological Charge of Thermodynamic Potentials}\label{Sec4}
%%%%%%%%%%%%%%%%%%%%%%%%%%%%%%%%%%%%%%%%%%%%%%%%%%%%
%%%%%%%%%%%%%%%%%%%%%%%%%%%%%%%%%%%%%%%%%%%%%%%%%%%%
One aspect of black hole studies that has garnered significant attention is the topological examination of their thermodynamic properties. This approach, based on Duan's topological current theory, allows us to investigate critical points associated with specific features of black holes, determining the stability or instability of particular points and enabling the topological classification of these black holes \cite{wei2020topological,hazarika2024thermodynamic,alipour2023topological}. In Section \ref{Sec3}, we explored the thermodynamic characteristics of rotating KK black hole. In this section, we aim to analyze Hawking temperature and the generalized free energy from a topological perspective.

We begin by examining the phase transition associated with the Hawking temperature. To this end, we define a thermodynamic potential using Eq. \eqref{Temp} in the following form \cite{wei2022topology}
\begin{eqnarray}
\Phi =\frac{1}{\sin \theta} T_+,
\end{eqnarray}
where $T_+$ is the Hawking temperature given by Eq. \eqref{Temp}. Two dimensional vector field of $\Phi$ can be represented in polar coordinates from the gradient of this potential as \cite{wu2024thermodynamical,bai2023topology,gogoi2023thermodynamic}
\begin{equation}
\phi^{\Phi}_{i}
=(\partial_{r_+}\Phi,\,\partial_{\theta}\Phi),\qquad i=r,\,\theta,
\end{equation}
normalized by the relation $n_i^{\Phi}=\phi^{\Phi}_i/||\phi^\Phi||$.
The zero points of this vector field, where the vectors converge or diverge, are located at $\phi^{\Phi} = (0,\,0)$, the angular component, $\phi^\theta$, vanishes at $\theta = \pi/2$, and the radial component of the critical points are determined by the condition $\partial_{r_+}\Phi\big|_{r_+=r_c}=0$. This condition is equivalent to $\partial_{r_+}T_H = 0$, which precisely matches the location of the second--order phase transition identified in Section \ref{Sec3}.

The topological charge $w_i$ of a zero point $r_{c_i}$ is determined by evaluating  the rotation of the vectors in a closed contour $c_i$ enclosing that point: If the rotation direction $\phi_r-\phi_\theta$ is clockwise, the topological number is $-1$; if counterclockwise, it is $+1$. The topological number for other points in the vector space is zero. The value of $w_i$ determines the topological nature of the critical point.

\begin{figure}[ht!]
%\centering
 %   \begin{subfigure}[b]{0.5\textwidth}
    \centering
  \includegraphics[width=5.7cm]{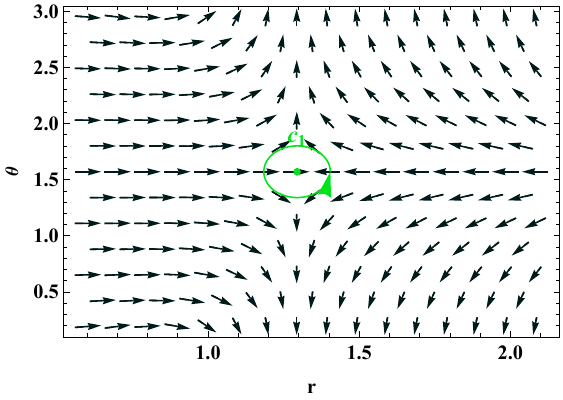} \hspace{0.5cm}
  \includegraphics[width=5.8cm]{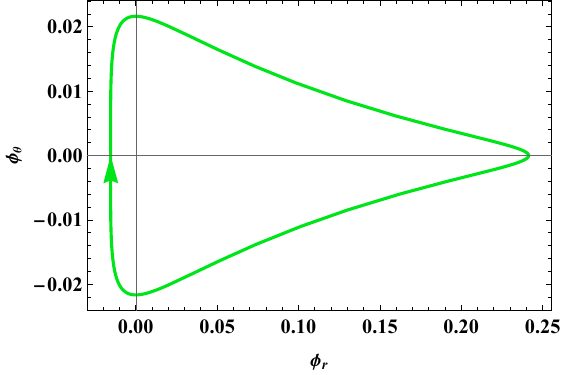} \hspace{-0.2cm}\\
%\caption{}
%    \end{subfigure}%
    \caption{Parameters are set as $\gamma=0.5$, $\lambda=10$, and $a=0.6$. Left panel: Vector space of potential $\Phi$ which contains a zero point at $r_{c_1}=1.2945$. Right panel: The variation of $\phi_\theta$ versus $\phi_r$ for zero point $r_{c_1}$. The direction of curves assigns a topological charge of $-1$.}\label{fig:TopoTemp}
\end{figure}

Figure \ref{fig:TopoTemp} displays the vector space for the potential $\Phi$ considering $\gamma=0.5$, $\lambda=10$, and $a=0.6$.
In this vector space, there exists a zero point at $r_{c_1}=1.2945$. By examining the rotation of $\phi_r-\phi_\theta$ curve, we find that the topological number of $r_{c_1}$ is $-1$. This topological number indicates the presence of a conventional critical point within the vector space of potential $\Phi$ \cite{wei2022topology}.

\begin{figure}[ht!]
\centering
 %   \begin{subfigure}[b]{0.5\textwidth}
 %   \centering
  \includegraphics[width=5.7cm]{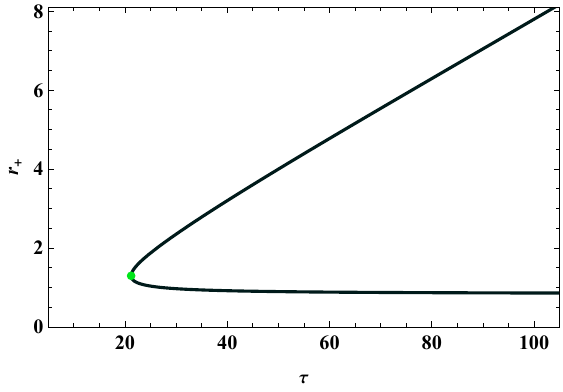} \hspace{-0.2cm}\\
%\caption{}
%    \end{subfigure}%
    \caption{$r_+-\tau$ curve considering $\gamma=0.5$, $\lambda=10$, and $a=0.6$. The curve has two branches for $\tau>21.2$.}\label{fig:TopoTau}
\end{figure}

\begin{figure}[ht!]
%\centering
 %   \begin{subfigure}[b]{0.5\textwidth}
    \centering
  \includegraphics[width=5.7cm]{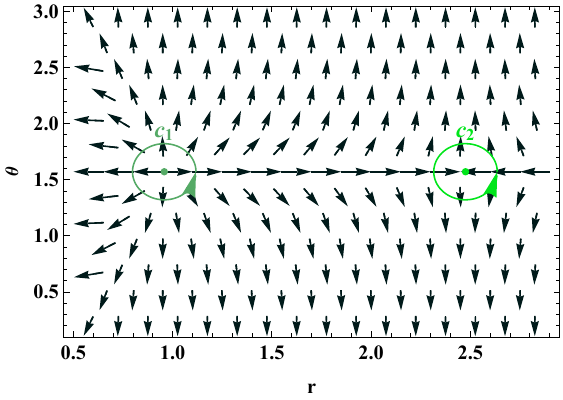} \hspace{0.5cm}
  \includegraphics[width=5.8cm]{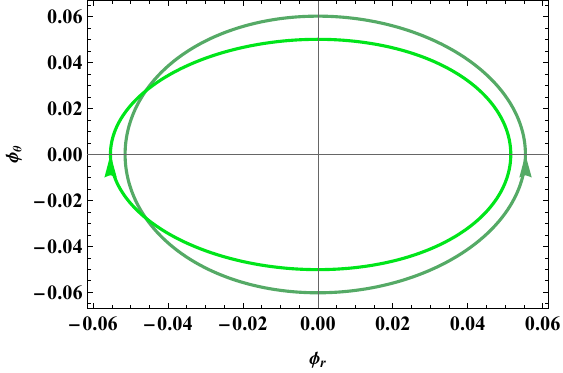} \hspace{-0.2cm}
%\caption{}
%    \end{subfigure}%
    \caption{$\gamma=0.5$, $\lambda=10$, $a=0.6$, and $\tau=10\pi$. Left panel: Vector space of potential $\mathcal{F}$ that has two zero points at $r_{c_1}=0.96$ and $r_{c_2}=2.48$. Right panel: Variation of $\phi_\theta$ as function of $\phi_r$ around zero points $r_{c_1}$ and $r_{c_2}$ that indicates the topological number of $+1$ and $-1$, respectively.}\label{fig:TopoHelm}
\end{figure}

To achieve a universal topological classification, we now aim to study the black hole using an off--shell thermodynamic potential. Following the framework of Wei et al. \cite{wei2022black}, we define the off--shell generalized free energy using Eqs. \eqref{Mplus} and \eqref{Ent} as
\begin{equation}
\begin{aligned}
\mathcal{F}=& M_+-\frac{S_+}{\tau}
\\
=& \frac{e^{r_+/\lambda } \left(\lambda  r_+ e^{r_+/\lambda }-\sqrt{\lambda  \left(\lambda  r_+^2 e^{\frac{2 r_+}{\lambda }}-\gamma  \left(a^2+r_+^2\right) (\lambda +r_+)\right)}\right)}{\gamma  (\lambda +r_+)}-\frac{\pi  \left(a^2+r_+^2\right)}{\tau }
\end{aligned}
\end{equation}
where $\tau$ is considered as the inverse temperature of the black hole. This potential can be represented in vector space using the vectors
\begin{align}
\phi^{\mathcal{F}}_{i}=(\partial_{r_+}\mathcal{F},\,-\cot \theta \csc \theta),\qquad i=r,\,\theta,
\end{align}
which are normalized similarly to the potential vectors $\phi_i^\Phi$. The zero point of this vector space is located at $\partial_{r_+}\mathcal{F}\big|_{r_+=r_c}=0$ and $\theta=\pi/2$. By solving $\partial_{r_+}\mathcal{F}=0$, we can derive a relation for $\tau$ in terms of the horizon radius, given by
\begin{footnotesize}
\begin{align}
\tau=\frac{4 \pi  \lambda  r_+ e^{\frac{-r_+}{\lambda }} (\lambda +r_+) \left(r_+ \left(-e^{\frac{r_+}{\lambda } } \sqrt{\lambda  \left(\lambda  r_+^2 e^{\frac{2 r_+}{\lambda }}-\gamma  \left(a^2+r_+^2\right) (\lambda +r_+)\right)}+\gamma  r_+^2+\lambda  r_+ \left(\gamma -e^{\frac{2 r_+}{\lambda }}\right)\right)+a^2 \gamma  (\lambda +r_+)\right)}{\lambda  e^{\frac{r_+}{\lambda } } \left(a^2 \left(2 \lambda ^2+2 r_+^2+3 \lambda  r_+\right)+r_+^3 (\lambda +2 r_+)\right)-\left(a^2 (\lambda +2 r_+)+r_+ \left(2 \lambda ^2+2 r_+^2+3 \lambda  r_+\right)\right) \sqrt{\lambda  \left(\lambda  r_+^2 e^{\frac{2 r_+}{\lambda }}-\gamma  \left(a^2+r_+^2\right) (\lambda +r_+)\right)}}.
\end{align}
\end{footnotesize}
Figure \ref{fig:TopoTau} illustrates the behavior of $r_+$ as a function of $\tau$ with the choice of $\gamma=0.5$, $\lambda=10$, and $a=0.6$.
This curve changes direction at $\tau_c=21.2$ and there are two possible solutions for $\tau>\tau_c$, one corresponding to a smaller black hole and one to a larger black hole. Figure \ref{fig:TopoHelm} displays the vector space of potential $\mathcal{F}$ for $\gamma=0.5$, $\lambda=10$, $a=0.6$, and $\tau=10\pi$.
In this vector space, there are two zero points at $r_{c_1}=0.96$ and $r_{c_2}=2.48$. By plotting the curve $\phi_r-\phi_\theta$, the corresponding topological numbers for these zero points are $+1$ and $-1$, respectively.

Finally, we classify the rotating KK black hole based on its total topological charges, $W_T = \sum_i w_i$. Based on the global properties of the vector field $\mathcal{F}$, black holes are categorized into four universal classes:
$W^{+1}$ and $W^{-1}$ classes, these represent systems where the total topological charge is non--zero ($W = +1$ or $W = -1$). These classes typically describe black holes that maintain a single stable or unstable branch across the parameter space without the annihilation of critical points, and $W^{0+}$ and $W^{0-}$ classes, these classes represent black holes with a total topological charge of zero. This occurs when the black hole has a pair of critical points, one Innermost and one Outermost, with opposite winding numbers ($w = +1$ and $w = -1$) that cancel each other out \cite{wei2024universal,zhu2025universal,chen2025universal}.
\\The distinction between $W^{0+}$ and $W^{0-}$ lies in the nature of their zero points. The Innermost critical point is at a smaller horizon radius, while the Outermost critical point is at a larger radius. In the $W^{0+}$ class, the Outermost critical point serves as a unstable defect with a negative topological number.

Our investigation shows that for the rotating KK black hole, there are two critical points whose total topological charge is $W = 0$. Since the topological defect at the larger horizon radius (outermost) has a negative topological number, we conclude that the rotating KK black hole belongs to the $W^{0+}$ universal class. This topological signature remains unchanged under variations in the spin and the KK parameters $\gamma$ and $\lambda$, demonstrating the fundamental stability of its thermodynamic structure.
%%%%%%%%%%%%%%%%%%%%%%%%%%%%%%%%%%%%%%%%%%%%%%%%%%%%
%%%%%%%%%%%%%%%%%%%%%%%%%%%%%%%%%%%%%%%%%%%%%%%%%%%%
\section{Optical Properties and shadow constraints}\label{Sec6}
%%%%%%%%%%%%%%%%%%%%%%%%%%%%%%%%%%%%%%%%%%%%%%%%%%%%
%%%%%%%%%%%%%%%%%%%%%%%%%%%%%%%%%%%%%%%%%%%%%%%%%%%%
The study of the Black Hole shadow provides a crucial method for exploring theories of gravity, particularly in the strong--field regime. The shadow emerges as the optical manifestation of strong gravitational lensing of light in the vicinity of a black hole, presenting as a two--dimensional dark region against a brighter background when viewed from a distance \cite{synge1966escape,cunningham1973optical,bardeen1973timelike,teo2003spherical,hackmann2010analytical,islam2024investigating}. To determine the contour of the shadow, we analyze the null geodesics in the spacetime.

\begin{figure}[ht!]
	%\centering
	%   \begin{subfigure}[b]{0.5\textwidth}
		\centering
		\includegraphics[width=5.7cm]{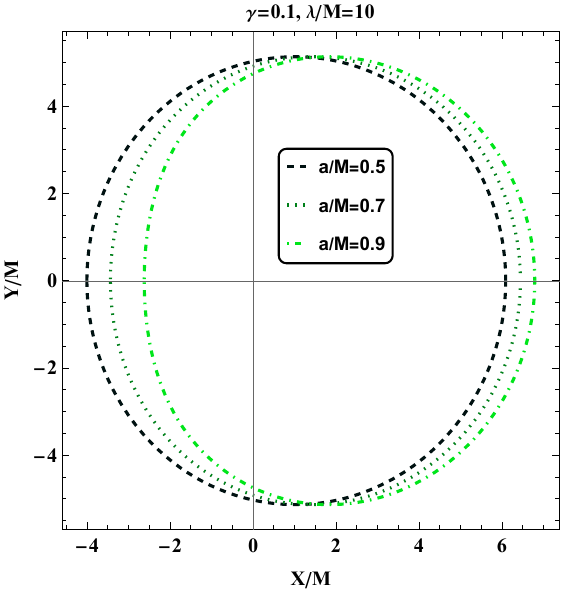} \hspace{0.2cm}
		\includegraphics[width=5.7cm]{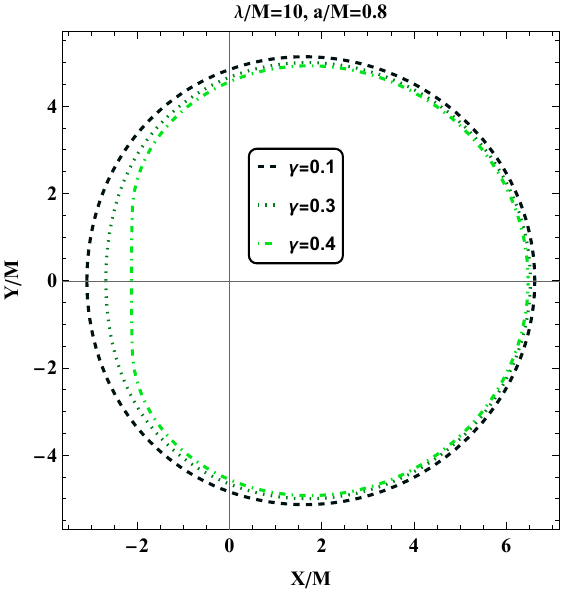} \hspace{0.2cm}
		\includegraphics[width=5.7cm]{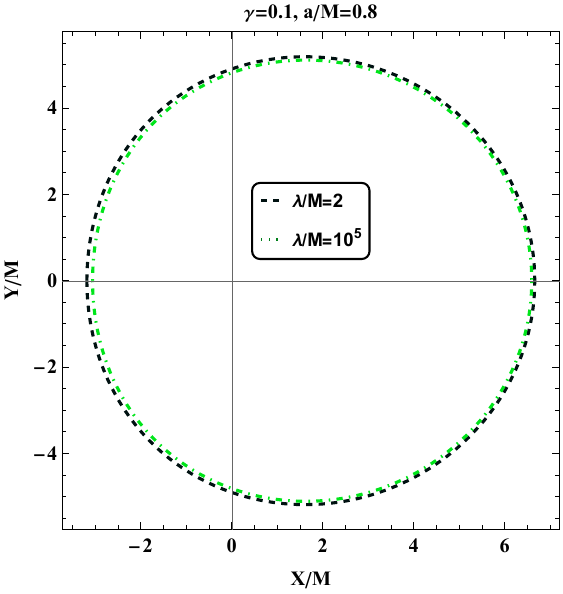} \hspace{-0.2cm}
		%\caption{}
		%    \end{subfigure}%
	\caption{The black hole shadow silhouettes under variations of the black hole parameters for an observer on equatorial plane.}\label{fig:Shadow}
\end{figure}

The geodesic equations of massless particles can be effectively solved using the Hamilton--Jacobi formalism. The Hamilton-Jacobi equation is given by \cite{carter1968global,chandrasekhar1998mathematical}
\begin{equation}
	\frac{\partial S}{\partial \delta} = -\frac{1}{2} g^{\mu\nu} \frac{\partial S}{\partial x^\mu} \frac{\partial S}{\partial x^\nu},
\end{equation}
where $\delta$ is the affine parameter along the geodesic. Due to the symmetries of the spacetime, the action $S$ for massless particles can be separated as \cite{chandrasekhar1998mathematical}
\begin{equation}\label{action}
	S =E t + L_z \phi + S_r(r) + S_\theta(\theta),
\end{equation}
where $S_r(r)$ and $S_\theta(\theta)$ are functions of the radial and polar coordinates, respectively. Also, the quantities $E$ and $L_z$ are the conserved quantities associated with the time-translation and rotational symmetries, defined as
\begin{align}
	E=-\frac{\partial S}{\partial t}=-g_{t\phi}\dot{\phi}-g_{tt}\dot{t},\qquad L_z=\frac{\partial S}{\partial\phi}=g_{\phi\phi}\dot{\phi}+g_{\phi t}\dot{t}.
\end{align}
The separability of Eq. \eqref{action} introduces a third conserved quantity, the Carter constant $K$. This separation leads to the following equations of motion for the null geodesics
\begin{equation}
\begin{aligned}
	\Sigma \dot{t} &= \frac{r^2+a^2}{\Delta} \left(E(r^2+a^2) - aL_z\right) - \frac{a}{\Delta} \left(L_z - aE\sin^2\theta\right), \\
	\Sigma \dot{\phi} &= \frac{a}{\Delta} \left(E(r^2+a^2) - aL_z\right) - \frac{1}{\Delta \sin^2\theta} \left(L_z - aE\sin^2\theta\right), \\
	\Sigma \dot{r} &= \sqrt{R(r)}, \\
	\Sigma \dot{\theta} &= \sqrt{\Theta(\theta)},
\end{aligned}
\end{equation}
where the radial and angular potentials are given by
%\begin{equation}
%	\begin{aligned}
%		R(r)\equiv&\Delta^2\left(\frac{\partial S_r}{\partial r}\right)^2,\\
%		\Theta(\theta)\equiv&\Delta^2\left(\frac{\partial S_\theta}{\partial \theta}\right)^2
%	\end{aligned}
%\end{equation}
\begin{equation}
\begin{aligned}\label{RTheta}
	R(r) &= \left(E(r^2+a^2) - aL_z\right)^2 - \Delta [K + (L_z - aE)^2],\\
	\Theta(\theta) &= K - \cot^2\theta L_z^2 - a^2 \cos^2\theta E^2.
\end{aligned}
\end{equation}
The boundary of the black hole shadow is determined by the set of unstable spherical photon orbits, which circle the black hole at a constant radius r = $r_{ph}$. The circular photon orbits are located at $R(r_{\rm ph})=0=R'(r_{\rm ph})$. 
\\Solving these equations allows us to determine the critical impact parameters of the photon orbits. We define two dimensionless impact parameters \cite{chandrasekhar1998mathematical}
\begin{equation}
	\xi = \frac{L_z}{E} \quad \text{and} \quad \eta = \frac{K}{E^2}.
\end{equation}
Solving the system of equations yields the critical values of these parameters, $\xi_c$ and $\eta_c$, as functions of the photon sphere radius $r_{\rm ph}$ as
\begin{equation}
	\begin{aligned}
		\xi_c=&-\frac{\gamma  M^2 \left(a^2 (\lambda +2 r_{\rm ph})+r_{\rm ph} \left(4 \lambda ^2+2 r_{\rm ph}^2+5 \lambda  r_{\rm ph}\right)\right)+2 \lambda ^2 e^{\frac{2 r_{\rm ph}}{\lambda }} \left(a^2 (M+r_{\rm ph})+r_{\rm ph}^2 (r_{\rm ph}-3 M)\right)}{2 a \lambda ^2 (r_{\rm ph}-M) e^{\frac{2 r_{\rm ph}}{\lambda }}-a \gamma  M^2 (\lambda +2 r_{\rm ph})}\\
		\eta_c=&
		\bigg(r_{\rm ph}^2 \left(-4 \gamma  \lambda ^2 M^2 e^{\frac{2 r_{\rm ph}}{\lambda }} \left(a^2 \left(4 \lambda ^2+4 r_{\rm ph}^2+6 \lambda  r_{\rm ph}\right)+r_{\rm ph} (r_{\rm ph}-3 M) \left(4 \lambda ^2+2 r_{\rm ph}^2+5 \lambda  r_{\rm ph}\right)\right.\right)
		\\&\left.-4 \lambda ^4 r_{\rm ph} e^{\frac{4 r_{\rm ph}}{\lambda }} \left(r_{\rm ph} (r_{\rm ph}-3 M)^2-4 a^2 M\right)-\gamma ^2 M^4 \left(4 \lambda ^2+2 r_{\rm ph}^2+5 \lambda  r_{\rm ph}\right)^2\right)\bigg)
		\\&\bigg/\left(a \gamma  M^2 (\lambda +2 r_{\rm ph})+2 a \lambda ^2 (M-r_{\rm ph}) e^{\frac{2 r_{\rm ph}}{\lambda }}\right)^2
	\end{aligned}
\end{equation}
For the photon sphere, the two largest solutions yield the retrograde radius, $r_{\rm ph}^+$, and the prograde radius, $r_{\rm ph}^-$.
To visualize the shadow, we project these critical orbits onto the sky of a distant observer. For an observer located in the equatorial plane $\theta_o=\pi/2$ and radial infinity, the celestial coordinates $(X,\,Y)$ are related to the critical impact parameters by
\begin{equation}
	\begin{aligned}
		X=&-\xi\csc\theta_o\\
		Y=&\pm\sqrt{\eta+a^2 \cos^2 \theta_o-\xi^2\cot^2 \theta_o}
	\end{aligned}
\end{equation}
By plotting $Y$ as a function of $X$ for all outermost unstable photon orbits, we trace the boundary of the black hole shadow. The contour plots in Figure \ref{fig:Shadow} illustrate the variation of the black hole shadow and show how its shape is affected by the black hole's own properties and the underlying gravitational theory.
As shown, enhancing the spin parameter, the shadow's shape becomes noticeably asymmetrical, causing its center to distort and shift. This happens because of frame--dragging, where the spinning black hole pulls spacetime along with it. This alters the paths of photons that form the shadow's edge. Photons rotating in the opposite direction are pushed farther out, while those rotating in the same direction can orbit closer. This difference creates the D--shaped distortion we observe, which gets stronger as the spin parameter increases. Also, increasing $\gamma$ and $\lambda$ makes the shadow smaller. This can be understood as the extra--dimensional vector field making the gravitational pull less strong. In this modified potential, the photon sphere becomes smaller and moves closer to the black hole. Also, it is observed that the shadow's sensitivity to changes in parameter $\lambda$ is considerably smaller than its sensitivity to the spin and the parameter $\gamma$.

To further quantify the visual results from the shadow contours, we first analyze the linear radius of the shadow, denoted by $R_{\rm sh}$. This quantity serves as an intuitive measure of the shadow's overall size and is obtained from
\begin{equation}
	R_{\rm sh}=\frac{(X_t-X_r)^2+Y_t^2}{2|X_r-X_t|},
\end{equation}
where the subscripts $t$ and $r$ represent the top and rightmost points of the shadow.
Due th the Event Horizon Telescope (EHT) data for Sgr A$^*$, the magnitude of shadow lies within the range $\{4.55\,M,\,5.22\,M\}$ and $\{4.21\,M,\,5.56\,M\}$ for the $1\sigma$ and $2\sigma$ regions, respectively \cite{vagnozzi2023horizon}.

\begin{figure}[ht!]
		\centering
		\includegraphics[width=5.7cm]{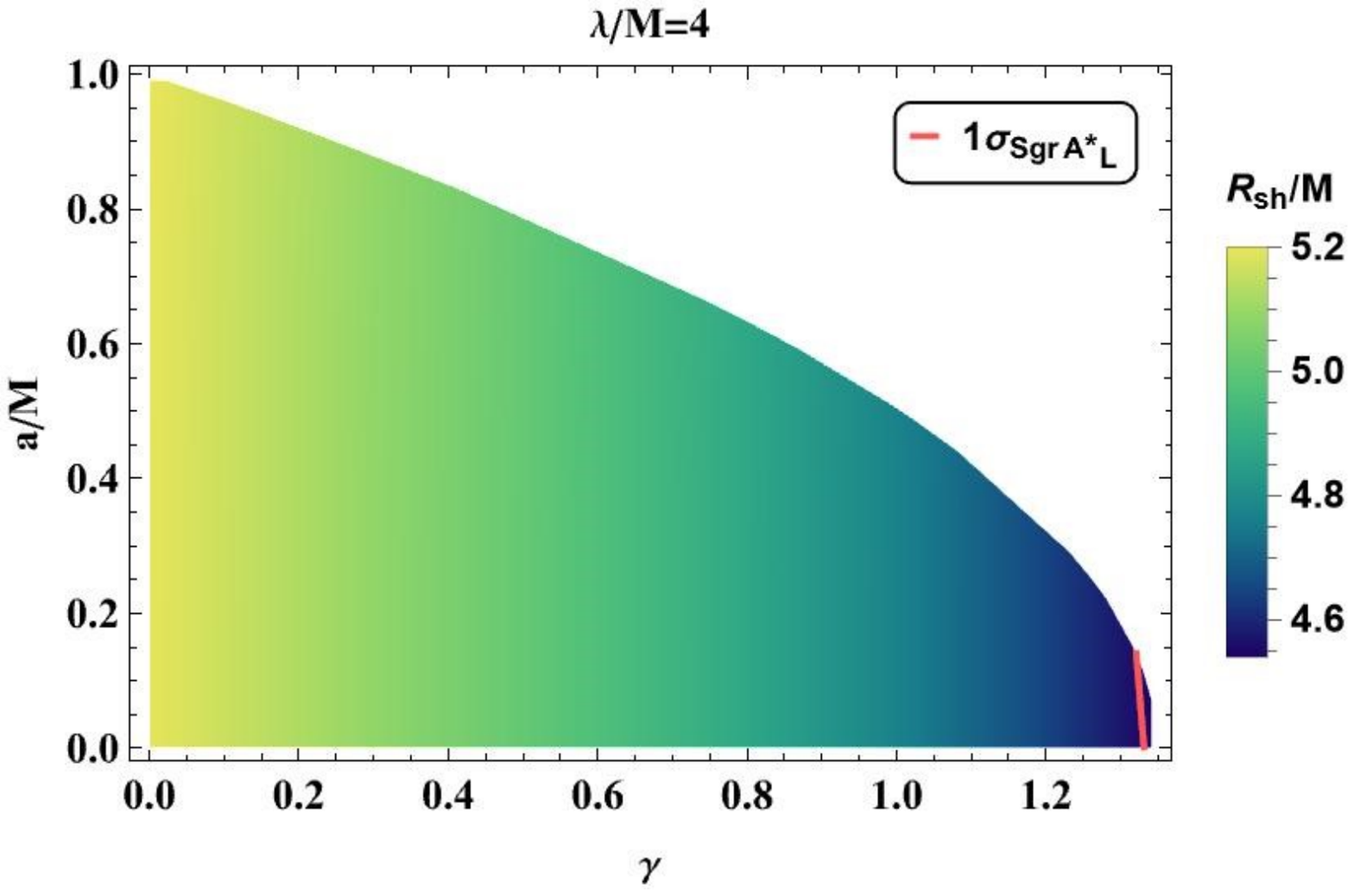} \hspace{0.2cm}
		\includegraphics[width=5.7cm]{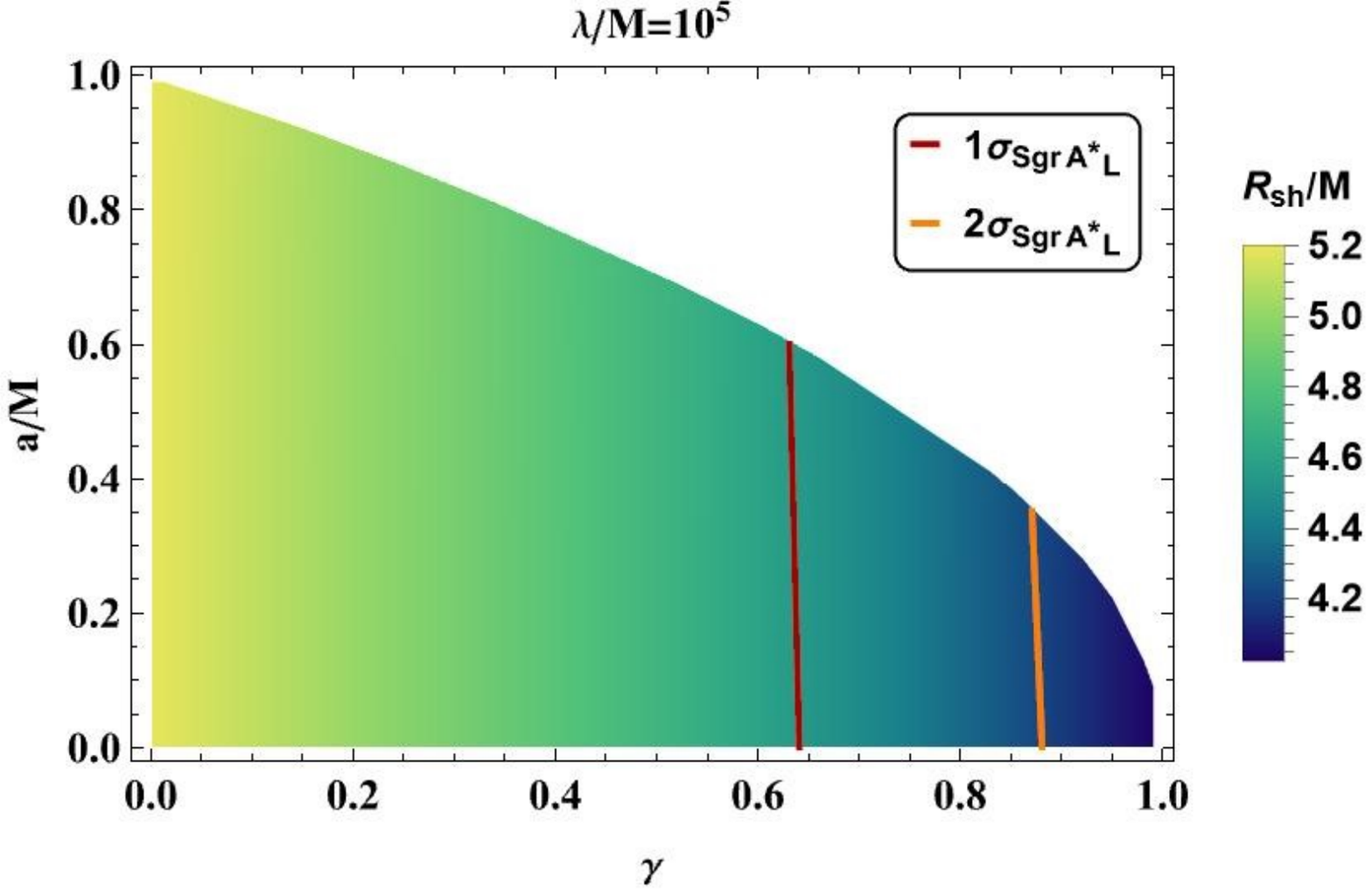} \hspace{-0.2cm}
	\caption{The variation of $R_sh/M$ as a function of $a/M$ and $\gamma$. Dashed lines represent the lower bounds of $1\sigma$ and $2\sigma$ regions for Sgr A$^*$ based on the data of EHT.}\label{fig:ShadowL}
\end{figure}

Figure \ref{fig:ShadowL} displays $Rs$ as a function of spin and the parameter $\gamma$.
Our results are consistent with previous study \cite{jusufi2025black}, which investigated the shadow of a KK black hole and noted that the effect of the $\lambda$ parameter on the shadow's size is minimal compared to that of $\gamma$. We observe a similar behavior here: increasing the parameter $\gamma$ significantly impacts the shadow size, causing it to decrease.
Also, keeping $a/M$ and $\gamma$ constant, enhancing $\lambda/M$ causes in a decreases in the magnitude of the linear radius of shadow.
In addition, by comparing shadow of the rotating KK black hole with EHT data for Sgr A$^*$, while the magnitude of shadow lies below the upper bound of $1\sigma$ and $2\sigma$ regions, the upper bounds of these regions apply some constraint on the rotating KK black hole parameters, which is shown in Figure \ref{fig:ShadowL}. Specifically, for $\lambda/M=4$, shadow for $\gamma\preceq 1.3$ lies within the $1\sigma$ region.

\begin{figure}[ht!]
		\centering
		\includegraphics[width=6.5cm]{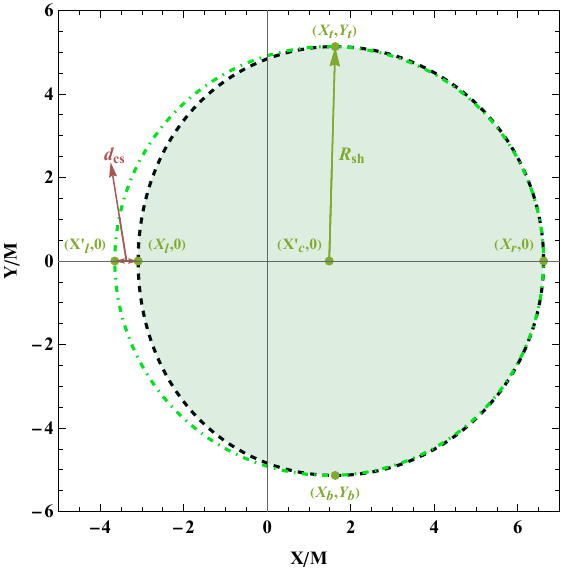} \hspace{-0.2cm}
	\caption{The shadow of a rotating KK black hole and its associated observables, considering $\gamma=0.1$, $\lambda/M=10$, and $a/M=0.8$. The subscripts $t, b, l,$ and $r$ denote the top, bottom, left, and rightmost points of the shadow, respectively. The superscript ${'}$ represents a perfect reference circle passing through the points $(X_r, 0)$, $(X_t, Y_t)$, and $(X_b, Y_b)$, and its center is indicated by the subscript $c$.}\label{fig:Shad_Obs}
\end{figure}

\begin{figure}[ht!]
	\centering
	\includegraphics[width=5.7cm]{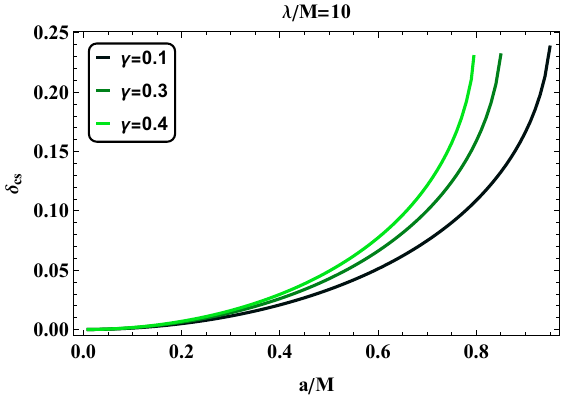} \hspace{0.2cm}
	\includegraphics[width=5.7cm]{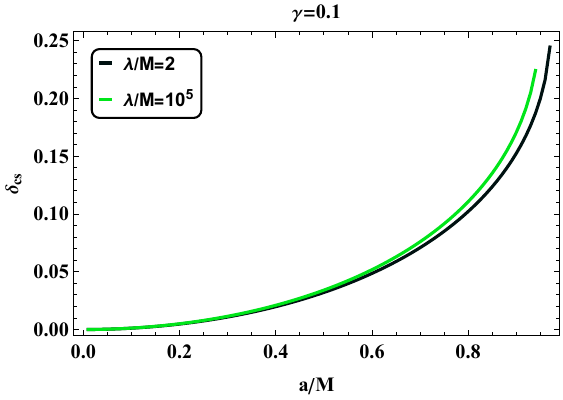} \hspace{-0.2cm}
	\caption{The distortion parameter as a function of the spin parameter. The plot shows results for several values of the KK parameters, while the observer's inclination angle is fixed at $\theta_o=\pi/2$.}\label{fig:ShadowDelta}
\end{figure}

Another key quantity we study in this section is the distortion parameter, which quantifies the deviation of the black hole shadow shape from a perfect circle.
This deformation is a consequence of the frame--dragging effect, and causes photons on retrograde orbits to require a larger impact parameter to be captured compared to those on prograde orbits.
Thus, the shadow appears compressed on one side, leading to an oblate shape. The distortion parameter is computed from
\begin{equation}
	\delta_{cs}=\frac{d_{cs}}{R_{\rm sh}},\qquad d_{cs}=X^{'}_l-X_l
\end{equation}
that the "prime" superscript denotes a perfect circle defined by the three points $(X_r, 0)$, $(X_t, Y_t)$, and $(X_b, Y_b)$. Furthermore, the four extremal points of the rotating KK black hole shadow are denoted by the subscripts $t$ (top), $b$ (bottom), $l$ (left), and $r$ (right), as depicted in Figure \ref{fig:Shad_Obs}.
The variations of $\delta_{cs}$ as a function of the spin parameter for various selections of KK parameters are illustrated in Figure \ref{fig:ShadowDelta}.
As observed, the distortion parameter shows an increasing trend with both the black hole dimensionless spin parameter $a/M$ and the KK parameters.
The dependence on $a/M$ is consistent with theoretical predictions for rotating black holes. An increase in the spin parameter indicates a faster rotating black hole, which increases the effect of frame--dragging. This phenomenon causes spacetime to be dragged more strongly around the black hole, leading to a greater compression of the shadow along the direction of rotation, thus increasing $\delta_{cs}$.
Furthermore, the results show that higher values of the KK parameters $\gamma$ and $\lambda/M$ also contribute to the increased distortion of the black hole shadow, increasing $\delta_{cs}$. Additionally, it can be concluded that, while the effect of varying spin parameter on the magnitude of $R_{\rm sh}$ is minimal, it significantly affects on the distortion parameter.

\begin{figure}[ht!]
	\centering
	\includegraphics[width=5.7cm]{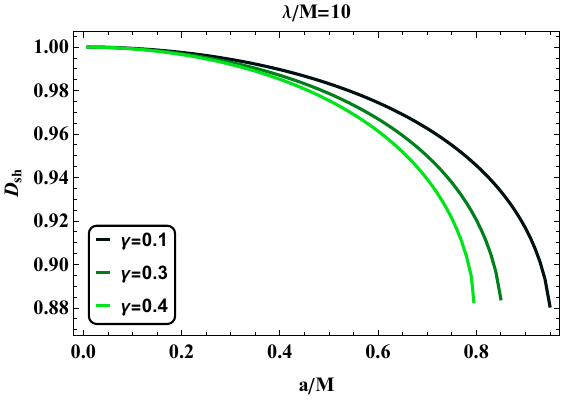} \hspace{0.2cm}
	\includegraphics[width=5.7cm]{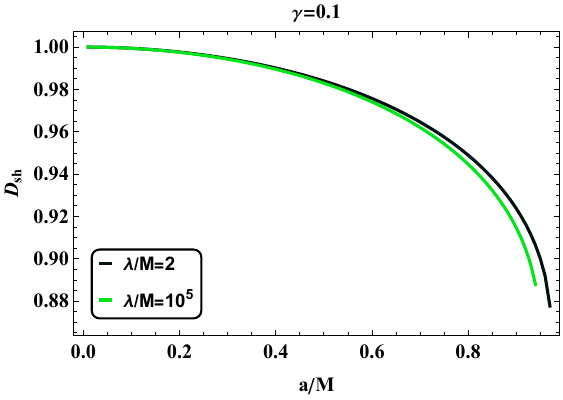} \hspace{-0.2cm}
	\caption{The variations of oblateness in terms of spin for some cases of KK parameters.}\label{fig:ShadowD}
\end{figure}

The deformation of the black hole shadow can be studied by the oblateness parameter which is defined as
\begin{equation}
	D_{\rm sh}=\frac{X_r-X_l}{Y_t-Y_b},
\end{equation}
that for the non--rotating black holes, $D_{\rm sh}$ becomes unity and is affected by the spin and other parameters of rotating black holes.
For the rotating KK black hole, deformation in terms of spin parameter and some cases of KK parameters is represented in Figure \ref{fig:ShadowD}.
As shown, for $a=0$, the oblateness is $D=1$. Also, in the regime of small $a/M$, varying KK parameters does not significantly affect $D_{\rm sh}$ but, by increasing $a/M$, enhancing KK parameters causes the black hole becomes more oblate.

\begin{figure}[ht!]
		\centering
		\includegraphics[width=5.7cm]{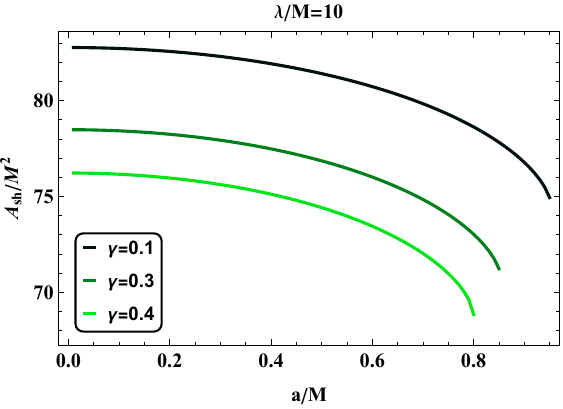} \hspace{0.2cm}
		\includegraphics[width=5.7cm]{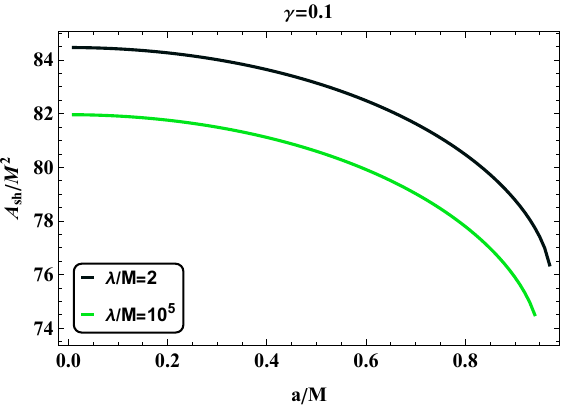} \hspace{-0.2cm}
	\caption{The shadow area as a function of $a/M$ for different values of KK parameters.}\label{fig:ShadowArea}
\end{figure}

To analyze the overall size of the black hole shadow, its area can also be calculated by integrating over the boundary of the shadow as
\begin{equation}
	A_{\rm sh}=2\int_{r_{\rm ph}^-}^{r_{\rm ph}^+}d r_{\rm ph}\big[ Y(r_{\rm ph})\partial_{r_{\rm ph}}X(r_{\rm ph})\big]
\end{equation}
The shadow area as a function of the spin parameter is illustrated in Figure \ref{fig:ShadowArea} for different values of the KK parameters.
As shown, the shadow area gets smaller with an increase in black hole spin. The shadow area also decreases when $\gamma$ or $\lambda$ increases. This decrease occurs because the extra-dimensional field weakens the gravitational pull, causing the photon sphere to shrink and move closer to the black hole.

%%%%%%%%%%%%%%%%%%%%%%%%%%%%%%%%%%%%%%%%%%%%%%%%%%%%
%%%%%%%%%%%%%%%%%%%%%%%%%%%%%%%%%%%%%%%%%%%%%%%%%%%%
\section{Luminosity of Accretion Disk}\label{Sec7}
%%%%%%%%%%%%%%%%%%%%%%%%%%%%%%%%%%%%%%%%%%%%%%%%%%%%
%%%%%%%%%%%%%%%%%%%%%%%%%%%%%%%%%%%%%%%%%%%%%%%%%%%%
Accretion Disks are structures surrounding black holes formed by diffusing gas and dust that  spirals inward.
In the static limit, Jusufi et al. \cite{jusufi2025black} showed that massive vector fields in KK gravity change the inner edge of the accretion disk. They found that increasing the KK parameters lowers the ISCO radius and raises the radiation efficiency.
Their results indicated that the black holes with greater KK parameters have higher maximum thermal profiles and maximum luminosity compared to ones with lower KK parameters.
In this section, we aim to study this features for a rotating black hole in the presence of KK gravity.
To model the disk, we use the Novikov--Thorne algorithm \cite{novikov1973astrophysics,page1974disk}, a relativistic geometric model for thin accretion disks in curved spacetime, that assumes the disk lies in the equatorial plane $\theta = \pi/2$ and that the particles follow nearly geodesic circular orbits, and the mass accretion rate $\dot{M}$ is constant over time, which we consider it as unity.

To study the motion of massive particles, Lagrangian method is employed as
\begin{equation}
	\mathcal{L}=\frac{1}{2}g_{\mu\nu}\dot{x}^{\mu}\dot{x}^{\nu}
\end{equation}
where $g_{\mu\nu}$ is the metric tensor and $\dot{x}$ represents the derivation of the coordinates $\mu\in(t,\,r,\,\phi)$. Using Euler--Lagrange equation and defining two conserved quantities $E$ and $L$, a marginally stable circular orbit is obtained from the following conditions
\begin{eqnarray}\label{rdot}
	\dot{r}=0\longrightarrow V_{\text{eff}}(r)=\frac{1}{2}E^2,\qquad\qquad\ddot{r}=0\longrightarrow \partial_r V_{\text{eff}}(r)=0.
\end{eqnarray}
in which $V_{\rm eff}$ refers to the effective potential of massive particles.
The angular velocity refers to the rate at which particles in the accretion disk rotate around the central object and influences the dynamics of the disk and is expressed as
\begin{align}\label{omega}
	\Omega(r)&=\frac{\sqrt{\left(\partial_r g_{t\phi}\right)^2-\partial_r g_{tt} \partial_r g_{\phi \phi}}-\partial_r g_{t\phi}}{\partial_r g_{\phi \phi}}.
\end{align}
The energy of a particle in the accretion disk and its angular momentum is obtained from 
\begin{equation}
	\label{Er}
	\begin{aligned}
		E(r)&=-\frac{g_{tt}+g_{t\phi} \Omega }{\sqrt{-g_{tt}-2 g_{t\phi} \Omega-g_{\phi\phi} \Omega^2}}
		,
	\end{aligned}
\end{equation}
and
\begin{equation}
	\label{Lr}
	\begin{aligned}
		L(r)&=
		\frac{g_{t\phi}+g_{\phi\phi} \Omega }{\sqrt{-g_{tt}-2 g_{t\phi} \Omega-g_{\phi\phi} \Omega^2}}
		.
	\end{aligned}
\end{equation}
The ISCO is the smallest radius at which a massive particle can orbit a central object in a stable manner and $\partial^2_r V_{\text{eff}}(r)\big|_{r=r_{\rm ISCO}}=0\rightarrow\partial_r L(r)\big|_{r=r_{\rm ISCO}}=0$ \cite{boshkayev2016motion}.
%It plays a crucial role in determining the inner edge of the accretion disk.

\begin{figure}[htbp!]
	%\centering
	%   \begin{subfigure}[b]{0.5\textwidth}
		\centering
		\includegraphics[width=5.7cm]{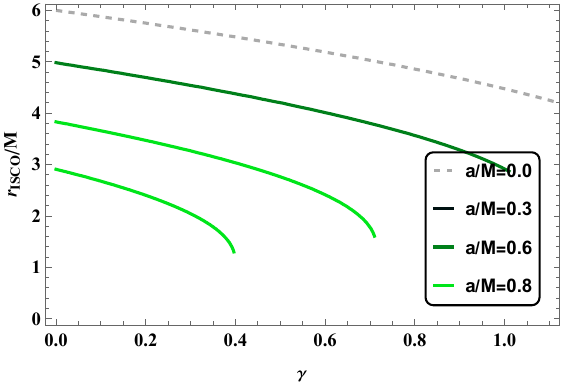} \hspace{0.5cm}
		\includegraphics[width=5.7cm]{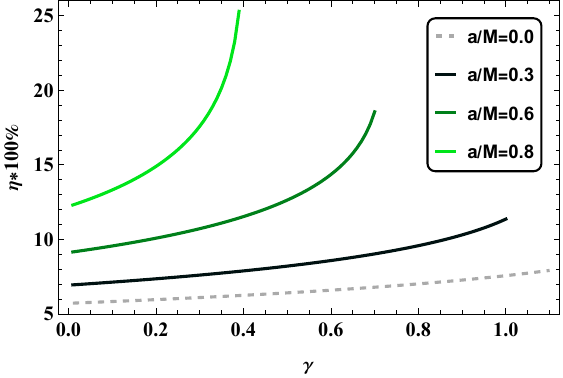} \hspace{-0.2cm}
		%\caption{}
		%    \end{subfigure}%
	\caption{
		Variation of the ISCO radius and the radiation efficiency for $\lambda/M = 10$. The plot highlights how the KK parameters shift the stable orbit boundary and the efficiency.}\label{fig:ISCO}
\end{figure}

The efficiency of an accretion disk is determined by the ratio of the energy converted into radiation to the mass accretion rate. It indicates how effectively the disk transforms gravitational energy into radiation and is computed from
\begin{eqnarray}
	\eta=1-E(r_{\rm ISCO}).
\end{eqnarray} 
We started our research by creating physical boundaries, which controlled the parameter values needed to form at least a black hole horizon. Our research about accretion disk dynamics, which shows that the KK framework depends on the spin parameter $a$, lies within this range. In Figure \ref{fig:ISCO}, the variations of the ISCO radius as a function of $\gamma$ changes are illustrated.
For $a/M = \gamma = 0$, the ISCO radius of the black hole under study is equal to the ISCO radius of the Schwarzschild black hole $6M$.
%An increase in the size of either parameter $a/M$ or $\gamma$ leads to a decrease in the ISCO radius.
We observed that for a fixed $\gamma$, increasing the spin significantly reduces the ISCO radius. Also, the affect of KK parameters become stronger when the spin value reaches $0.8$. This means rotation makes KK parameters more influential.
Additionally, in this figure, the efficiency for different values of $a/M$ is plotted against changes in the $\gamma$ parameter. In the case of $a/M = \gamma = 0$, the efficiency of the black hole equals $5.71$, which is the same as the efficiency of the Schwarzschild black hole, and increasing the size of either parameter $a/M$ or $\gamma$ increases the efficiency. Also, the highest growth rate observed in the rapidly rotating regime (higher $a/M$).

\begin{figure}[htbp!]
	%\centering
	%   \begin{subfigure}[b]{0.5\textwidth}
		\centering
		\includegraphics[width=5.7cm]{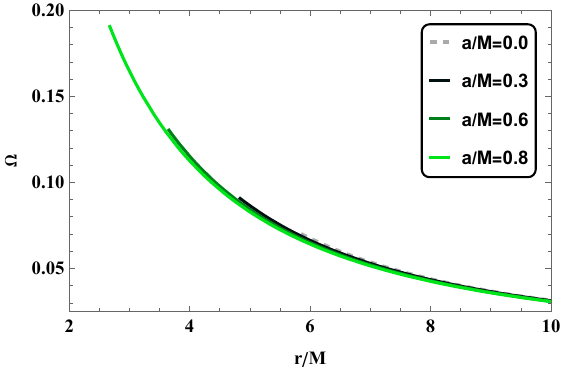} \hspace{0.2cm}
		\includegraphics[width=5.7cm]{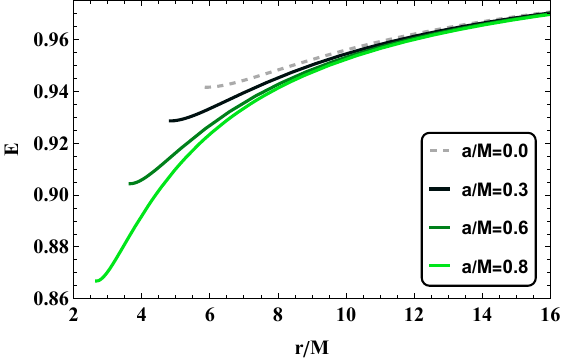} \hspace{0.2cm}
		\includegraphics[width=5.7cm]{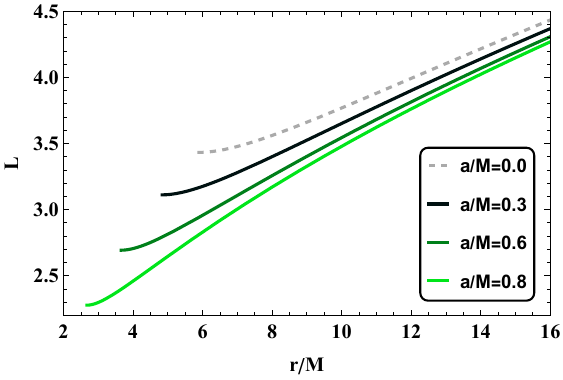} \hspace{-0.2cm}
		%\caption{}
		%    \end{subfigure}%
	\caption{The behavior of the angular velocity, specific energy, and angular momentum as a function of the radial coordinate for $\gamma = 0.1$ and $\lambda/M = 10$. The curves illustrate the influence of spin and extra--dimensional fields on the orbital dynamics.}\label{fig:OmEL}
\end{figure}

Furthermore, the variations of $\Omega$, $E$, and $L$ for a test particle for different values of $a$ are illustrated in Figure \ref{fig:OmEL}.
As $r/M$ increases, $\Omega$ exhibits a decreasing behavior. This study showed that when we increased parameter $a/M$, the disk particles started to move slower in their circular paths.
On the other hand, $E$ and $L$ shows an increasing trend. Additionally, higher values of $a/M$ lead to a decrease in both the specific energy and angular momentum of the orbiting particles, especially in the region near the ISCO (small $r/M$). Furthermore, at large distances, the energy converges to unity as expected.

\begin{figure}[htbp!]
	%\centering
	%   \begin{subfigure}[b]{0.5\textwidth}
		\centering
		\includegraphics[width=5.7cm]{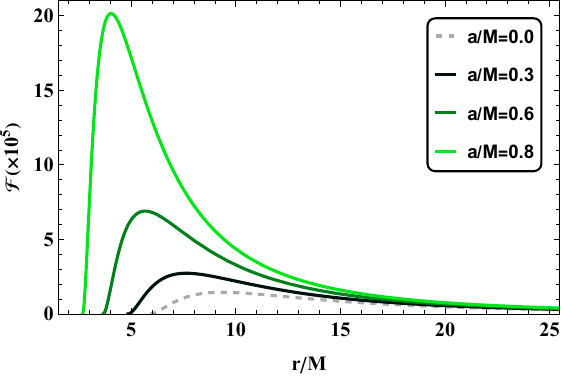} \hspace{0.5cm}
		\includegraphics[width=5.7cm]{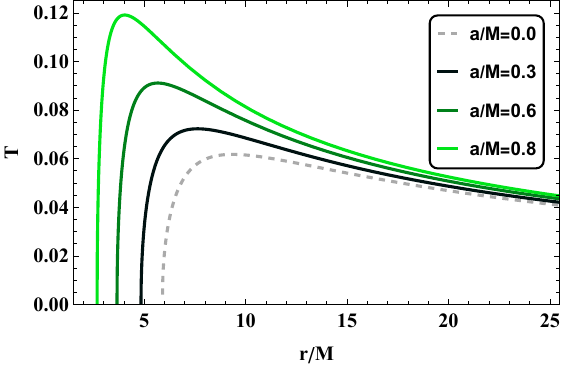} \hspace{-0.2cm}
		%\caption{}
		%    \end{subfigure}%
	\caption{The energy flux and the radiation temperature of the thin accretion disk as a function of radius for $\lambda/M = 10$ and $\gamma = 0.1$. The peaks indicate the region of maximum thermal radiation.}\label{fig:FT}
\end{figure}

Another quantity we study in this section is the energy flux, that represents the energy emitted from the accretion disk and is dependent on the radial distance and the properties of the disk. This quantity, employing Eqs. \eqref{omega} to \eqref{Lr} is obtained from \cite{novikov1973astrophysics,page1974disk}
\begin{eqnarray}\label{flux}
	\mathfrak{F}(r)=-\frac{1}{4\pi\sqrt{-g}}\frac{\partial_r \Omega(r)}{(E(r)-\Omega(r) L(r))^2}\int_{r_{\text{ISCO}}}^{r}(E(\tilde{r})-\Omega(\tilde{r}) L(\tilde{r}))\partial_{\tilde{r}}L(\tilde{r}) d \tilde{r},
\end{eqnarray}
where $\sqrt{-g}=\sqrt{-g_{rr}(g_{tt}g_{\phi\phi}-g^2_{t\phi})}$.
Assuming the disk is in local thermodynamic equilibrium, it radiates as a perfect blackbody, and the radiation temperature $T(r)$ relates to the distribution of energy in the disk through the Stefan--Boltzmann law
\begin{eqnarray}
	\mathfrak{F}(r)=\sigma_{\rm SB}T^4(r),
\end{eqnarray}
where $\sigma_{\rm SB}$ indicates the Stefan--Boltzmann constant. By studying the temperature distribution, we can explore how the extra--dimensional fields and the black hole's spin affect the thermal intensity and overall luminosity of the accretion process.
In Figure \ref{fig:FT}, the variations of energy flux and radiation temperature for different values of $a/M$ are represented.
At small $r/M$, the trend of changes in both quantities, the energy flux and radiation temperature, is sharply increasing, reaching a maximum value, and then, with an increase in $r/M$, the trend becomes decreasing. It is also observed that for $a/M = 0$, the maximum energy flux and temperature radiation are smaller in size and occur at a larger radius. However, with the increase in $a/M$, this maximum shifts to a smaller radius and has a larger magnitude. This indicates that rotation makes the disk hotter, brighter, and more compact.

\begin{figure}[htbp!]
	%\centering
	%   \begin{subfigure}[b]{0.5\textwidth}
		\centering
		\includegraphics[width=7.5cm]{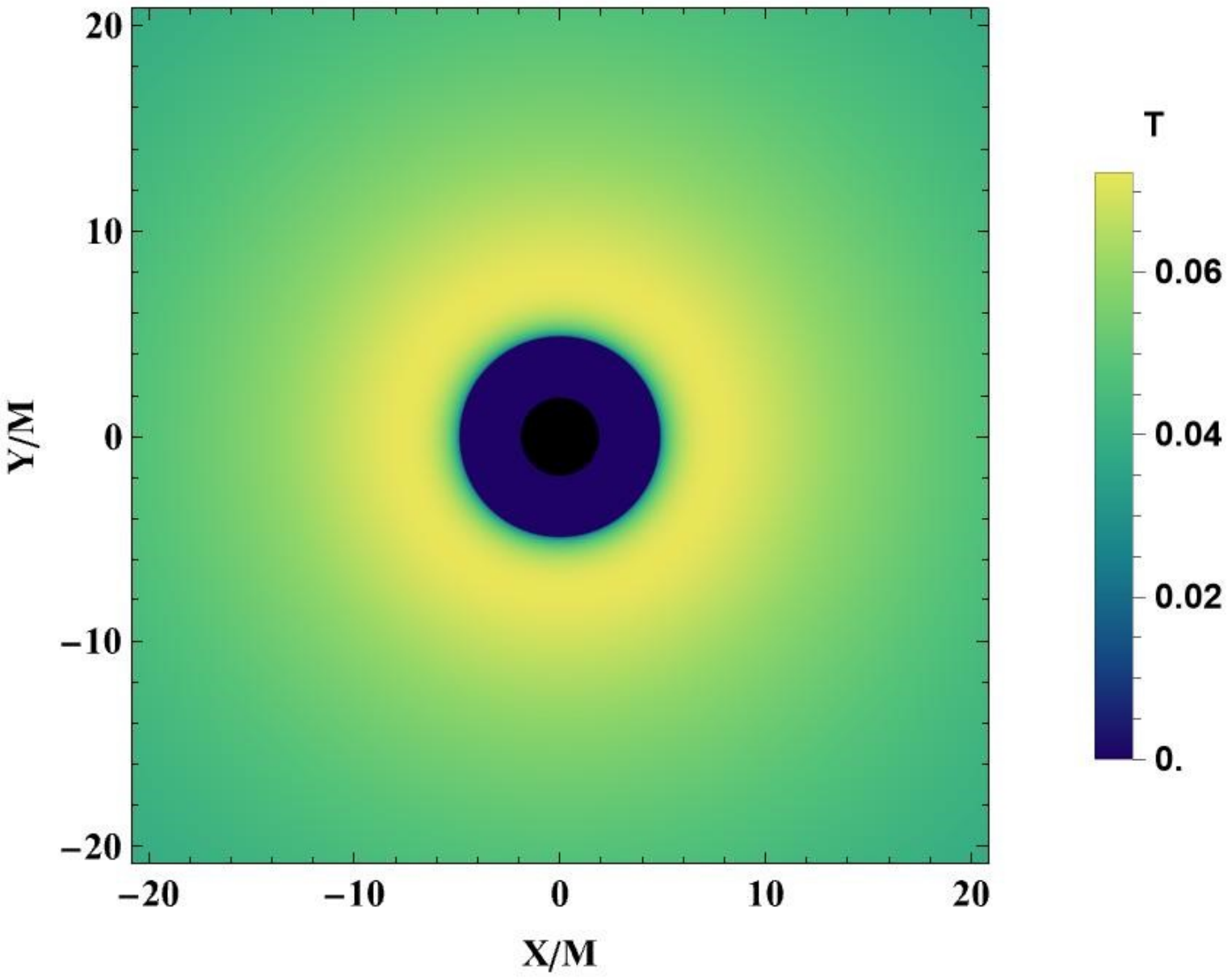} \hspace{0.5cm}
		\includegraphics[width=7.cm]{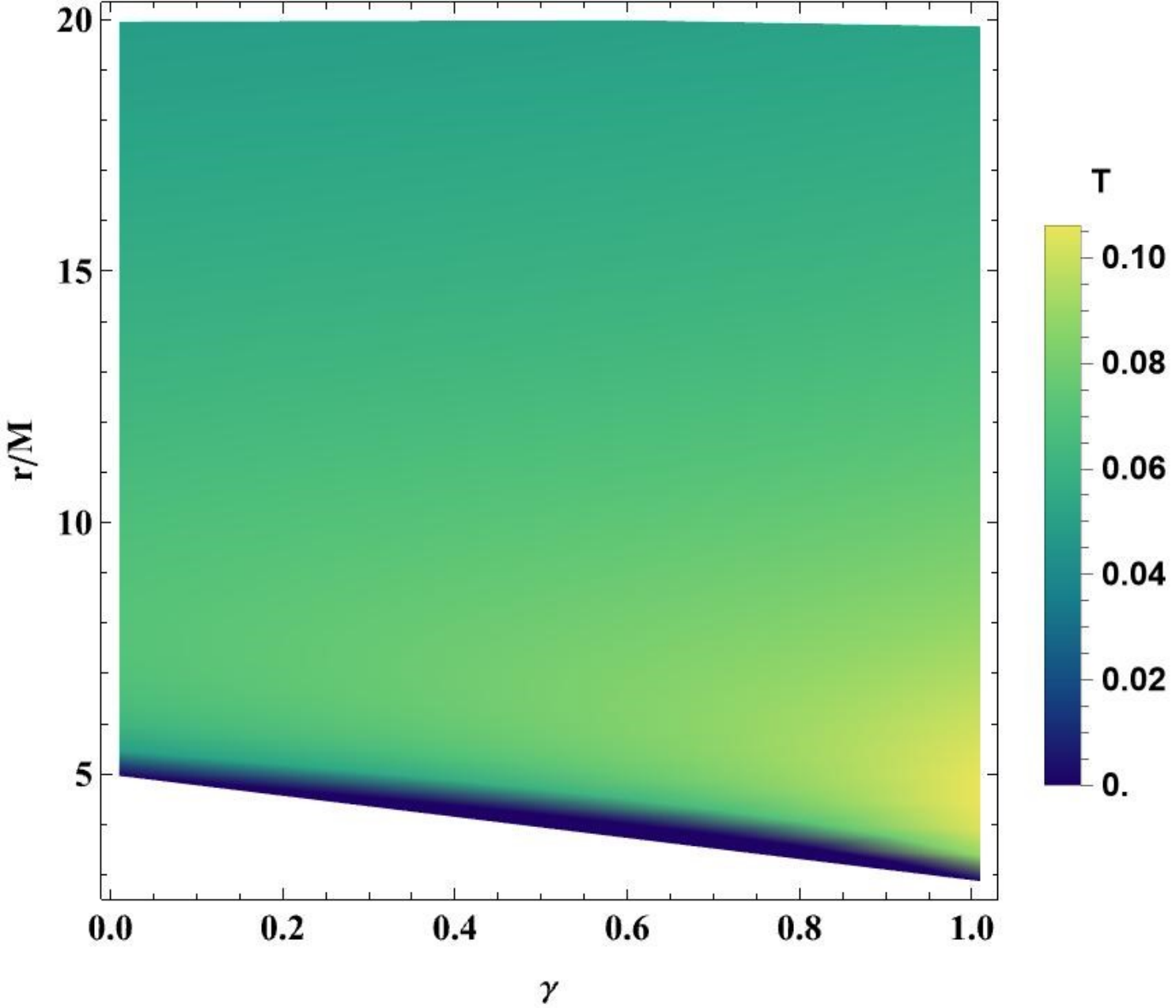} \hspace{-0.2cm}
		%\caption{}
		%    \end{subfigure}%
	\caption{Density profile of the radiation temperature considering $\lambda/M=10$ and $a/M=0.3$, in the X--Y lane for $\gamma=0.1$ in left panel, and by varying $r/M$ and $\gamma$ in right panel}\label{fig:DensT}
\end{figure}

The density of radiation temperature variations is plotted in Figure \ref{fig:DensT}.
In the left panel, these changes in the $X-Y$ plane are shown for $\gamma = 0.1$. The boundary of the black disk and the colored area is the horizon of the black hole. From $r_h$ to $r_{\rm ISCO}$, the temperature radiation is zero, and as $r/M$ increases, as seen in Figure \ref{fig:FT}, $T$ initially reaches its maximum value and then decreases. In the right panel, the changes in radiation temperature with varying $r/M$ and $\gamma$ are observed. It is clear that an increase in $\gamma$ leads to a larger maximum temperature radiation, with the maximum occurring at a smaller radius compared to cases with smaller $\gamma$ (as studied in Ref. \cite{jusufi2025black}).

In addition, the differential luminosity describes the energy reaching an observer at infinity, normalized to the radial distance from the central object. It is calculated based on the energy flux and helps quantify the total radiation emitted by the disk and reads \cite{novikov1973astrophysics,page1974disk,bambi2012code}
\begin{eqnarray}
	\frac{d \mathfrak{L}_\infty}{d \ln r}=4\pi r \sqrt{-g} E(r) \mathfrak{F}(r).
\end{eqnarray}

Furthermore, the last quantity we examine  in this section is the spectral luminosity, which refers to the distribution of luminosity as a function of frequency and is obtained from \cite{boshkayev2020accretion,boshkayev2021luminosity}
\begin{eqnarray}
	\nu\mathfrak{L}_{\nu,\infty}=\frac{15}{\pi^4}\int_{r_i}^\infty \left(\frac{d \mathfrak{L}_\infty}{d \ln r}\right)\frac{\left(u(r)y\right)^4}{M^2 \mathfrak{F}(r)}\frac{1}{\exp\left(\dfrac{u(r)y}{\left[M^2 \mathfrak{F}(r)\right]^{1/4}}\right)-1}d\ln r.
\end{eqnarray}

\begin{figure}[htbp!]
	%\centering
	%   \begin{subfigure}[b]{0.5\textwidth}
		\centering
		\includegraphics[width=5.7cm]{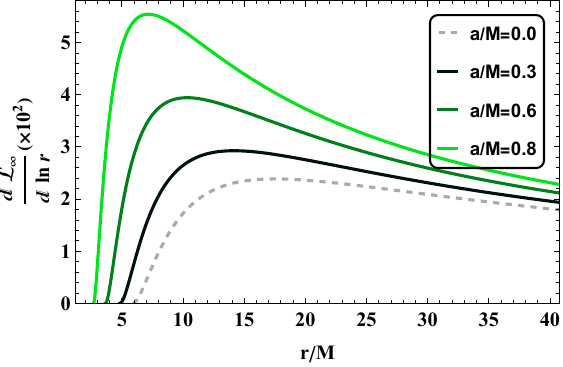} \hspace{0.5cm}
		\includegraphics[width=5.6cm]{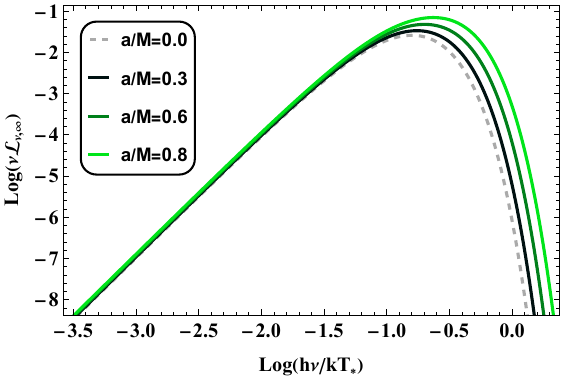} \hspace{-0.2cm}
		%\caption{}
		%    \end{subfigure}%
	\caption{The differential luminosity in terms of $r/M$ (left panel) and spectral luminosity as a function of frequency for $\lambda/M = 10$ and $\gamma = 0.1$ (right panel).}\label{fig:Lum}
\end{figure}

In Figure \ref{fig:Lum}, the variations of differential luminosity and spectral luminosity are illustrated.
It is observed that the changes in differential luminosity follow a similar trend to energy flux, meaning it initially reaches a maximum and then decreases, with an increase in $a/M$ having a similar effect on these variations. Additionally, it is noted that the magnitude of spectral luminosity increases with frequency, reaching its maximum, and then follows a decreasing trend. Moreover, at a constant frequency, an increase in $a/M$ results in an increase in the magnitude of spectral luminosity, raises its peak and shifts it toward higher frequency.
%%%%%%%%%%%%%%%%%%%%%%%%%%%%%%%%%%%%%%%%%%%%%%%%%%%%
%%%%%%%%%%%%%%%%%%%%%%%%%%%%%%%%%%%%%%%%%%%%%%%%%%%%
%\hspace{8.5cm}
\section{Conclusion}\label{Sec12}
%%%%%%%%%%%%%%%%%%%%%%%%%%%%%%%%%%%%%%%%%%%%%%%%%%%%
%%%%%%%%%%%%%%%%%%%%%%%%%%%%%%%%%%%%%%%%%%%%%%%%%%%%
In this study, we analyze the horizon structure and thermodynamic stability of rotating KK black holes. By investigating the metric function, we mapped the allowed parameter space. We identified the critical points where the system changes from a black hole with two horizons to an extremal state, and then to a naked singularity. Our results show that the spin and the parameter $\gamma$ have a significant impact on the horizon radii, while the parameter $\lambda$ has a non--linear, saturating effect.

The study of ergosphere indicates that it is strongly affected by the spin parameter, which enlarges and deforms it through frame--dragging. Among the KK parameters, $\gamma$ has a noticeable influence on the ergoregion shape, while $\lambda$ plays a much weaker role.

The thermodynamic analysis shows that the Hawking temperature has a distinct peak, indicating a phase transition. We found that while increasing the spin or the parameter $\gamma$ causes a cooling effect at fixed radii, raising $\lambda$ makes the maximum temperature greater, however enhancing all spin and KK parameters shifts the thermal maximum to larger scales. This pattern is further supported by the heat capacity and generalized free energy. The phase transition points and energy minima move toward larger horizon radii as $a$ or KK parameters increases. Additionally, the presence of a black hole remnant, which is larger due to rotation and extra-dimensional fields, suggests that these parameters prevent complete evaporation.
%This could lead to stable astrophysical relics.

A key focus of this study is the topological classification of the thermodynamics of the rotating KK black hole. By examining the Hawking temperature through its topological aspects, we found a topological charge of $-1$. This confirms the existence of a conventional critical point in the system. To better understand the global thermodynamic structure, we examined the off--shell generalized free energy. Our results show that the black hole falls into the $W^{0+}$ universal class. This class has two critical points, whose total topological charges offset each other.

Furthermore, we investigate the shadow properties of the black hole in the presence of spin and KK parameters. We find that increasing the spin parameter significantly distorts the shadow, making it more asymmetric and more oblate, due to stronger frame dragging. The KK parameters generally reduce the shadow size and increase its distortion, although the effect of $\lambda$ is weaker than that of the spin and $\gamma$. Moreover, by comparing the EHT data for Sgr A$^*$ with the shadow of the rotating black hole in the presence of KK parameters, we determine the allowed parameters ranges for which the shadow lies within the permitted $1\sigma$ and $2\sigma$ regions.

In the study of accretion disk dynamics, our findings show that introducing the spin parameter to the KK framework greatly improves the energetic properties of the disk. We demonstrated that rotation creates a more compact and efficient accretion system. The reduction in the ISCO radius, especially at high spin rates, allows the disk to extend deeper into the strong--field region. This leads to a significant increase in radiative flux, radiation temperature, and both differential and spectral luminosities. Specifically, the inward shift of the luminosity peaks acts as a key signature, suggesting that rotating black holes in the presence of extra--dimensional fields are much more luminous and energetically active than their static counterparts.
%%%%%%%%%%%%%%%%%%%%%%%%%%%%%%%%%%%%%%%%%%%%%%%%%%%%
%%%%%%%%%%%%%%%%%%%%%%%%%%%%%%%%%%%%%%%%%%%%%%%%%%%%
\hspace{4cm}
\section*{Acknowledgements}
F. S. and H. H. are grateful to Excellence project FoS UHK 2203/2025-2026 for the financial support.
%%%%%%%%%%%%%%%%%%%%%%%%%%%%%%%%%%%%%%%%%%%%%%%%%%%%
%%%%%%%%%%%%%%%%%%%%%%%%%%%%%%%%%%%%%%%%%%%%%%%%%%%%
%\hspace{0.5cm}
\nocite{*}
\bibliographystyle{plain}

\end{document}